\documentclass[useAMS,usenatbib]{mn2e}
\usepackage{graphicx}    
\usepackage{dcolumn}     
\usepackage{bm}          
\usepackage{amssymb,amsmath}
\usepackage{caption}
\usepackage[labelfont=bf]{caption}
  
\usepackage{color}
\title[DNF - Galaxy photometric redshift by Directional Neighbourhood Fitting]
      {DNF - Galaxy photometric redshift by Directional Neighbourhood Fitting}
\author[J. De Vicente, E. S\'anchez and I. Sevilla-Noarbe]{J. De Vicente, 
 E. S\'anchez and I. Sevilla-Noarbe\thanks{E-mail:juan.vicente@ciemat.es, eusebio.sanchez@ciemat.es, ignacio.sevilla@ciemat.es},  \\
 Centro de Investigaciones Energ\'eticas, Medioambientales y Tecnol\'ogicas (CIEMAT),\\
 Avda. Complutense 40, E-28040, Madrid, Spain\\
}
\begin{document}

\date{\today}
\pagerange{1--11} \pubyear{2015}
\maketitle

\begin{abstract}
Wide field images taken in several photometric bands allow 
simultaneous measurement of redshifts for thousands of galaxies. 
 A variety of algorithms to make this measurement have appeared in
the last few years, the majority of which
can be classified as either template or training
based methods. Among the latter, nearest neighbour estimators
stand out as one of the most successful, in terms of
both precision and the quality of error estimation. In this paper we
describe the Directional Neighbourhood Fitting (DNF) algorithm based on the following: a new neighbourhood metric 
(Directional Neighbourhood), a photo-z estimation strategy (Neighbourhood
Fitting) and a method for generating the photo-z probability distribution function. We compare DNF with other well-known 
empirical photometric redshift tools using different public datasets (Sloan Digital Sky Survey, VIMOS VLT Deep Survey and Photo-z Accuracy Testing).
DNF achieves high-quality results with reliable error.
\end{abstract}

\section{Introduction}
\label{intro}

Current cosmological surveys require that the redshifts
of a large number of galaxies be measured to test and validate
different cosmological models. Spectroscopy provides high
resolution redshift measurement, at the cost of long
exposure times using large telescopes. Conversely, photometric 
redshift estimation (photo-z) is a powerful alternative in which individual precision is 
relaxed in favour of statistics.  In this case, a large number of low resolution
 galaxy redshifts are obtained from imaging in different bands. Several current and
future surveys are based on the measurement of massive
amounts of objects and rely on photometric redshifts to
accomplish their goals in the field of cosmology, e.g., Dark Energy Survey
(DES; \cite{Brenna2005}) and Large Synoptic Survey Telescope (LSST; \cite{lsstSRD}).

There are two basic approaches to photometric redshifts:
template-based methods and training-based methods. Template
based methods rely on a set of rest-frame spectra that are
redshifted to fit the galaxy photometric magnitudes
(\cite{1999MNRAS.310..540A}, \cite{2000ApJ...536..571B}, \cite{Hiperz2000}, \cite{Zebra2006}).

On the other hand, training based
methods (\cite{Connolly1995}, \cite{2004PASP..116..345C}, \cite{ArborZ2010}, \cite{TPZ2013},
 \cite{Holey2015}, \cite{Rau2015}) require a training sample, i.e. a set of imaged galaxies whose 
spectroscopic redshifts are known. Based on such a training sample, these methods are able to predict
the photometric redshifts for a larger galaxy sample
 without spectroscopic measurements.  A
recent review of the field of photometric redshifts can be
found in \cite{2012SPIE.8451E..34Z} and a comparison among
different methods in Photo-z Accuracy Testing (PHAT; \cite{PHAT}) and DES (\cite{Sanchez2014}).

The k-nearest neighbour (kNN) algorithm (\cite{Altman:1992:IKN}) is one of the
best-known training-based approaches.  For instance, \cite{2003AJ....125..580C} describe the application of this technique
to the Sloan Digital Sky Survey (SDSS) Early Data Release. The authors use
local interpolation in regions of colour space to reduce the
photo-z dispersion.  In \cite{Csabai2}, the authors instead
use a kd-tree search method to efficiently find the nearest
neighbours, followed by polynomial fitting.
\cite{2008ApJ...683...12B} apply kNN to generate full
photometric redshift probability density functions (PDFs)
for objects in SDSS Data Release 5, obtaining results
consistent with other machine learning
techniques. In another study, \cite{Wang2010} compared two
closely related methods, the kernel regression and the
nearest neighbours algorithms, for photometric redshift
estimation of quasar samples. Here, kernel regression
delivered more accurate predictions, whereas kNN showed its
superiority for high redshift
quasars. \cite{1538-3881-146-2-22} also estimate the
photometric redshifts of quasars to study the influence of
$k$ value and different input patterns on performance, while 
\cite{2008ApJ...689..709O} use the Nearest Neighbour algorithm
 to estimate the photo-z error (nearest neighbour error)
 of other photo-z techniques such as neural networks.

Another goal of photo-z algorithms is to estimate the
underlying redshift distribution $N(z)$ of galaxy samples.
In \cite{2008MNRAS.390..118L}, a nearest neighbour
approximation is used to estimate $N(z)$. The method does
not rely on photometric redshift estimates for individual
galaxies, which typically suffer from biases. Instead, it
assigns weights to galaxies in a spectroscopic subsample
such that the weighted distributions of photometric
observables match the corresponding distributions of the
photometric sample. The weights are estimated using a
nearest neighbour technique. The authors found that 'the weighting method
accurately recovers the underlying redshift distribution,
typically better than the photo-z reconstruction, provided
that the spectroscopic subsample spans the range of photometric
observables covered by the sample'. In \cite{2012ApJS..201...32S}, the nearest
neighbour weighting algorithm is used to calculate individual
redshift probability distributions $P(z)$ for galaxies in the
SDSS Data Release 8.

In this work, along with the more common Euclidean distance, we 
define and explore two new neighbourhood metrics:
the angular neighbourhood and the directional neighbourhood. Using
each of these three metrics, we implement two different
nearest neighbours approaches: nearest neighbours (kNN) and
neighbourhood fitting (NF). The different combinations of
algorithms/neighbourhoods are compared.  Among them,
Directional Neighbourhood Fitting (DNF) stands out as the
best performant, and is thus used as one of the photo-z estimators
for DES.

The rest of the paper is organized as follows: in
Section~\ref{sec:nneighbourhood} we introduce the Nearest Neighbour algorithm in the context of photometric
redshift estimation and define two new neighbourhoods to
address the problem. In Section~\ref{sec:testbench}, the SDSS
galaxy samples are introduced along
with the metrics involved in the comparison. In
Section~\ref{sec:nncomparison}, we describe kNN,
the most common nearest neighbours approach, and compare it to the new
neighbourhoods we have defined. In Section~\ref{sec:dnf}, the
DNF method is presented and
compared with previous approaches. In Section~\ref{sec:othersurveys},
we extend our analysis of DNF to the VIMOS VLT Deep Survey (VVDS) and PHAT datasets, which although smaller samples 
they provide interesting higher redshift galaxies.

Finally, the conclusions are
presented in Section~\ref{sec:conclusions}.

\section{Photometric redshift and neighbourhood}
\label{sec:nneighbourhood}

kNN is an algorithm commonly used for
pattern recognition and regression. kNN allows the
user to estimate an unknown feature of a specific
sample based on its closeness (in the space of observables) to
other training samples for which the feature is known. Common
kNN algorithms use the Euclidean distance defined in the
space of observables as a measurement of closeness or
neighbourhood.  In this section, we define two new
neighbourhoods that may provide advantages over the Euclidean
metric in some areas.  

In the photometric redshift estimation problem, the magnitudes of the galaxies 
in the set of filters constitute the space of observables. 
The so-called photometric sample is the set of galaxies represented
by points in this space (i.e. galaxies whose magnitudes have been measured) without
measured spectra. On the other hand, the training sample is constituted by the 
set of galaxies for which both magnitudes and spectra have been measured. 
In order to determine the redshift of a galaxy in the photometric sample, the kNN algorithm
 uses the training sample as a reference. The photo-z of the target galaxy is obtained from the
redshift of training galaxies that are nearby in this
multi-magnitude space. The definition of the neighbourhood,
the number of near training galaxies considered, and how
their redshifts are combined to compute the final photo-z 
are key elements in ensuring the accuracy of the estimation. The rest
of this section focuses on the first of these elements, the
neighbourhood.

The redshifts of the training sample form a hypersurface
defined over the multi-magnitude space. Given a point in this space, i.e. the galaxy whose redshift is
unknown, the surrounding redshift hypersurface may be abrupt
or smooth depending on the direction selected. The best
local hypersurface fitting will be achieved when the
neighbourhood is distributed along a smooth path on the
hypersurface. Since the Euclidean neighbourhood treats 
all directions equally, it may not be the optimal neighbourhood
for fitting in many cases. In this section, we explore other
neighbourhoods that provide smoother paths.

\begin{figure*} 
  \centering
  \leavevmode
  \begin{tabular}{ccc}
    \includegraphics[width=0.33\textwidth]{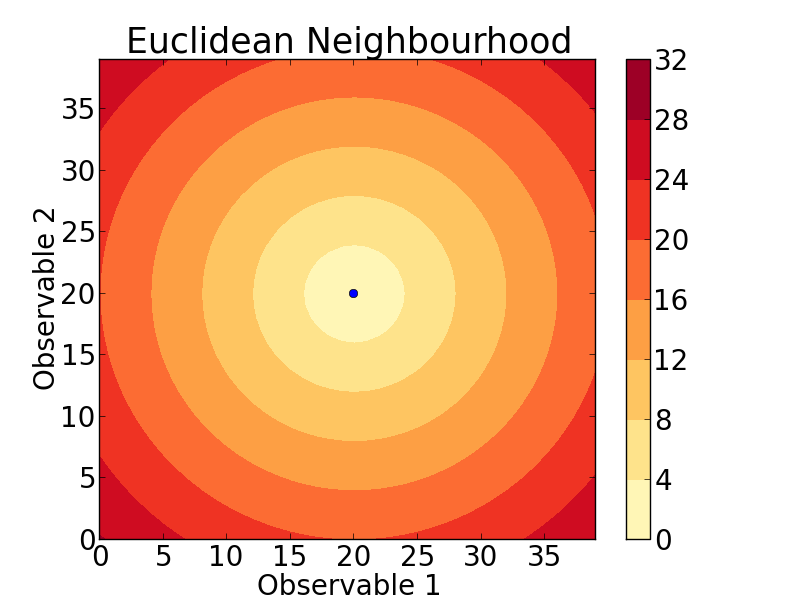} &
    \includegraphics[width=0.33\textwidth]{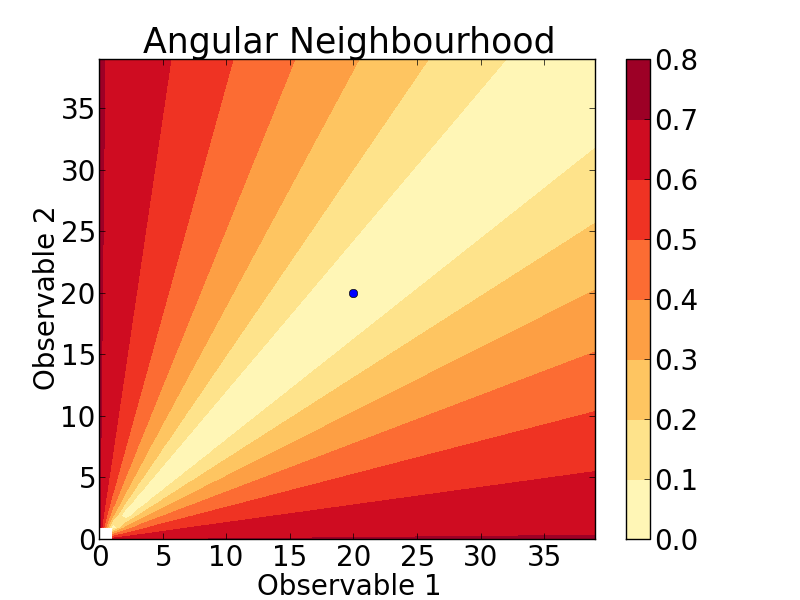} &
    \includegraphics[width=0.33\textwidth]{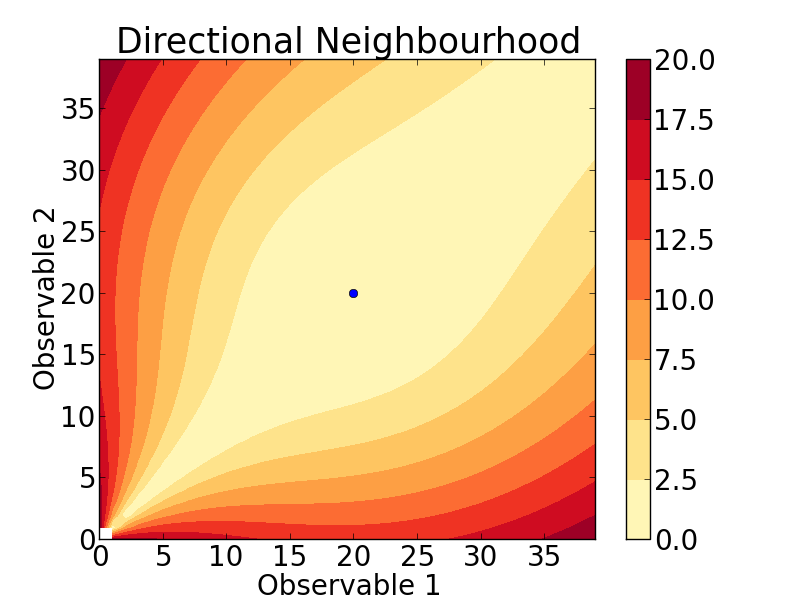}  \\
  \end{tabular}
  \captionsetup{labelfont=bf} \captionof{figure}{Neighbourhoods in two-dimensional space: Euclidean, angular and directional. 
The colour coding represents the 
distance to the point (20,20) in the space of observables. Unlike the Euclidean neighbourhood, angular and directional neighbourhoods 
account for the relative content of observables.}
  \label{fig:neighbourhood}
\end{figure*} 

To illustrate the definitions of the different neighbourhoods and for representativeness we consider a toy example of the space of 
observables limited to two dimensions. Fig.~\ref{fig:neighbourhood} shows the map of distances corresponding 
to a point of coordinates (20, 20) in a space of arbitrary observables. 
Euclidean is the
    common distance applied in kNN algorithms. Angular and
    directional are the new neighbourhoods described below that
consider galaxies as neighbours taken into account the relative content of the observables,
    in addition to the absolute one.

\subsection{Euclidean Neighbourhood}

The Euclidean neighbourhood (Fig.~\ref{fig:neighbourhood}) defined in the space of observables 
is the common distance metric used in kNN. The basic idea
behind kNN is that closeness among objects in the space of
observables implies closeness in terms of other non-observable
features. In photometric redshift prediction, the
observables are magnitudes or photon fluxes from
galaxies. kNN computes the photo-zs of a photometric sample
taking a neighbouring spectroscopic training sample as a
reference. Thus, the redshift of a galaxy in the photometric
sample is computed from the redshifts of its $k$ nearest
neighbours in the training sample. In the following
sections, we will use \textbf{e-kNN} to refer
to Euclidean nearest neighbours. The Euclidean neighbourhood, i.e., the distance between any pair of galaxies in the space of 
observables, can be quantified by equation ~\ref{eq:euclidean_dis}.

\begin{eqnarray}
  \label{eq:euclidean_dis}
   D=\sqrt{\sum_{i=i}^n (m^t_i-m^p_i)^2}
\end{eqnarray}

\noindent
where \textbf{$m^t$} and \textbf{$m^p$} are the
multi-magnitude vectors of training and photometric galaxies
respectively, and $n$ is the number of bands.

\subsection{Angular Neighbourhood}

Euclidean distance ensures that closely related galaxies in
a multi-magnitude space are assigned a similar
redshift. However, the Euclidean distance may be improved in
some aspects. Note that the Euclidean metric does not count as
neighbours galaxies with the same relative
multi-magnitude content (e.g. similar colour), but which are separated in
overall magnitude. There may be cases of galaxies of the
same type, at the same redshift, that differ in size and hence in apparent
magnitude. Thus, in spite of having the same redshift they would not be considered as neighbours.

In this work we have considered other metrics to mitigate
this drawback of Euclidean neighbourhoods. The first metric
explored is the angular Neighbourhood (\textbf{a-kNN}; Fig.~\ref{fig:neighbourhood}),
which is based on the normalized inner product (NIP; equation ~\ref{eq:nip}). NIP (\cite{Sanchez2014}) corresponds to
the cosine of the angle $\alpha$ formed by two
multi-magnitude vectors. The NIP metric considers two galaxies
as neighbours, hence having a close redshift, when they have
similar relative multi-magnitude content (e.g. similar colour) rather
than absolute multi-magnitude closeness. The metric is based
on the inner product definition

\begin{eqnarray}
\label{eq:nip}
NIP \equiv \cos\alpha = \frac{\sum_{i=i}^n m^t_i \cdot m^p_i}{\sum_{i=i}^n (m^t_i)^2\cdot \sum_{i=i}^n (m^p_i)^2}
\end{eqnarray}

Higher closeness corresponds to multi-magnitude
vectors in the same direction ($\alpha=0$). In this case
proportions among observables of two different objects are
maintained, and hence they are neighbours. As shown below,
 a-kNN improves e-kNN in terms of photo-z
dispersion.

\subsection{Directional neighbourhood}

In spite of the lower scatter of a-kNN compared to e-kNN for a few
neighbours, the performance of the angular neighbourhood degrades quickly 
when larger neighbourhoods are
considered, since it extends in a
divergent manner from the target point. To include the benefits of both approaches
we have defined the directional neighbourhood (DN; Fig.~\ref{fig:neighbourhood}) as the
product of Euclidean and angular neighbourhoods:
\begin{eqnarray}
\label{eq:dn}
DN=D^{2} \sin^{2}\alpha
\end{eqnarray}

\noindent
where $D$ is the Euclidean distance (equation ~\ref{eq:euclidean_dis}) and $\alpha$ is the angle
between the two multi-magnitude vectors defined by equation ~\ref{eq:nip}. 

As shown below (Section \ref{sec:dnf}), the benefits of directional
neighbourhood appear not just in terms of the nearest neighbour (\textbf{d-kNN}) but in
neighbourhood fitting, such that larger neighbourhoods can
be considered. In some sense, the Directional metric also accounts for 
kinship since the relative content of observables (i.e., proportionality) is maintained.

\section{Galaxy testbench and metrics}
\label{sec:testbench}

\subsection{Galaxy training and photometric samples}

To compare the performance of nearest neighbour photo-zs using
 different neighbourhood metrics,  we selected galaxies
from SDSS Data Release 10 (\cite{sdssDR10}). A million
objects with known photometry in the SDSS standard \textit{ugriz}
filters and with known spectroscopic redshift were selected. The
redshift range was $0.1< z < 0.7$. The basic SQL query used
for object selection is shown in Table
~\ref{table:SQLquery}.

\begin{table}
\caption{Galaxy selection SQL query}
\begin{tabular}{l}
   SELECT top number of objects \\
    \ \ $modelMag_u$, $modelMag_g$, $modelMag_r$, $modelMag_i$, \\
    \ \ $modelMag_z$,$modelMagErr_u$, $modelMagErr_g$, \\
    \  \ $modelMagErr_r$, $modelMagErr_i$,$modelMagErr_z$, \\
    \ \ z, objid \\
   FROM SpecPhoto WHERE z between 0.1 and 0.7 \\
\end{tabular}
\label{table:SQLquery}
\end{table} 

After excluding a few dozen objects tagged with erroneous magnitudes (magerr=9999), the sample was
shuffled and divided into two subsamples:
5,000 galaxies as the training sample and the rest as the photometric sample. Note that 5,000 galaxies
 is enough for training according to our prior tests since it is representative of the photometric sample.

\begin{figure*} 
  \centering
  \leavevmode
  \begin{tabular}{ccc}
    \includegraphics[width=0.33\textwidth]{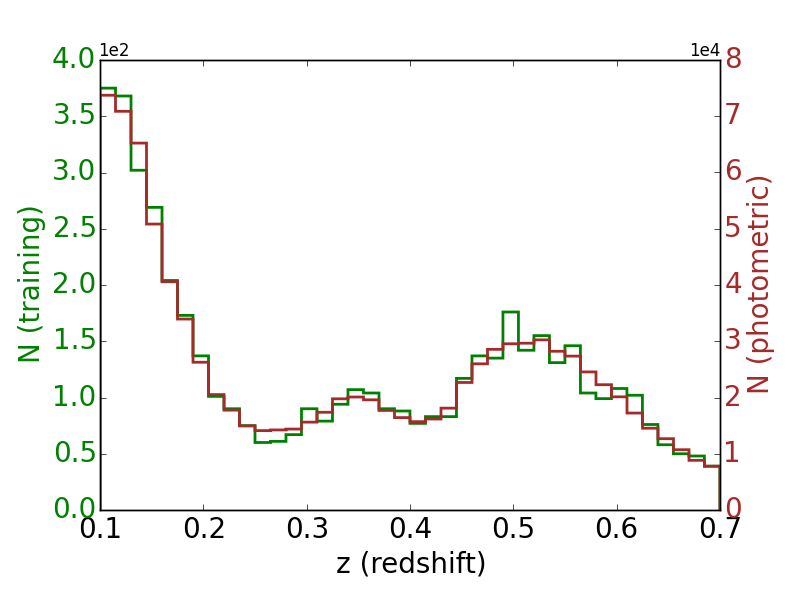} &
    \includegraphics[width=0.33\textwidth]{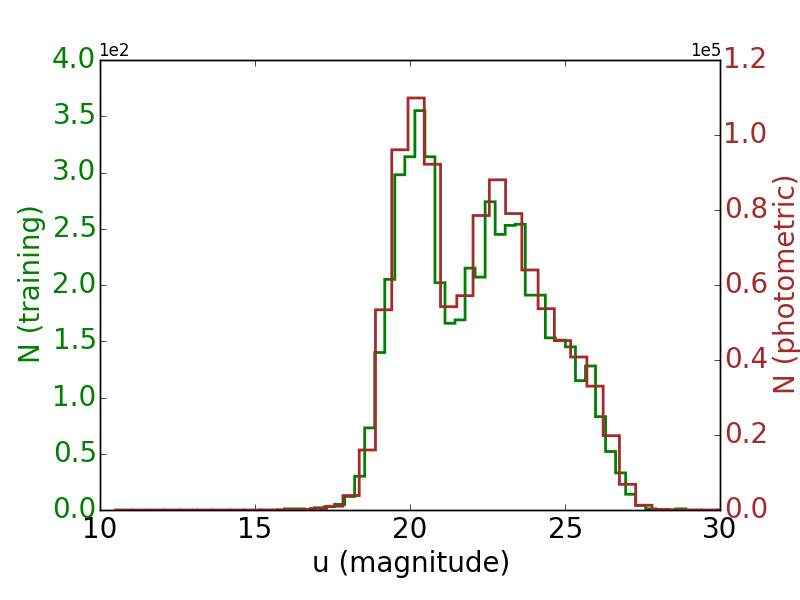} &
    \includegraphics[width=0.33\textwidth]{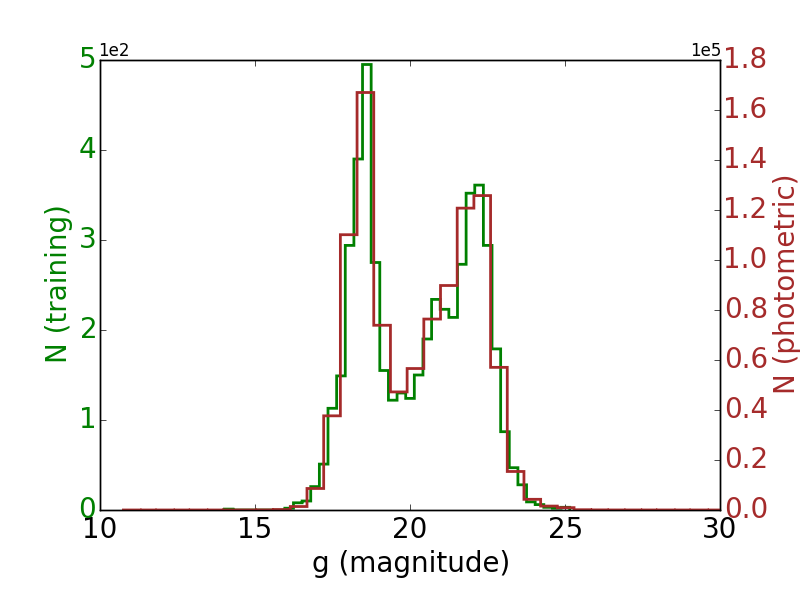} \\
    \includegraphics[width=0.33\textwidth]{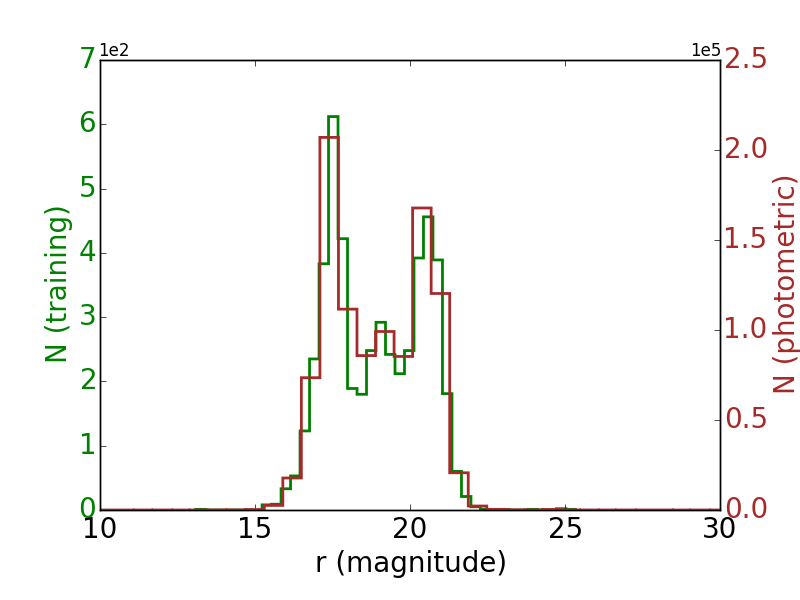} &
    \includegraphics[width=0.33\textwidth]{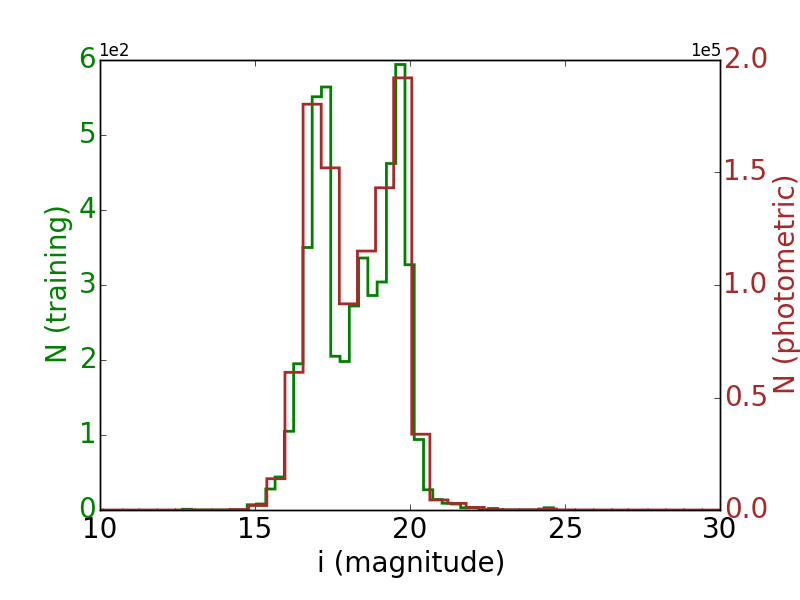} &
    \includegraphics[width=0.33\textwidth]{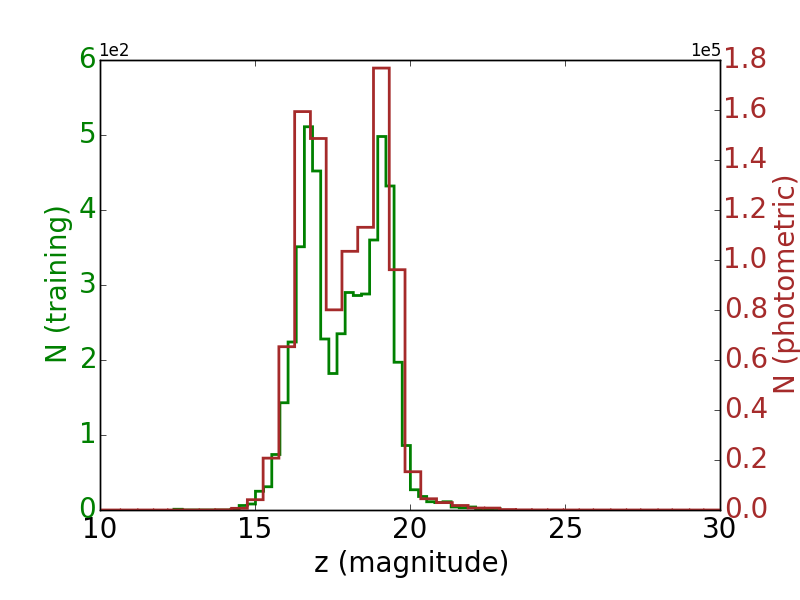} \\
  \end{tabular}
  \captionsetup{labelfont=bf} \captionof{figure}{True redshift
    and magnitude distributions of the training and photometric
    samples of the SDSS dataset used in this work. Note that in each plot the distributions are fully representative of one another.}
  \label{fig:zmagdistribution}
\end{figure*} 

Fig.~\ref{fig:zmagdistribution} shows the distribution in
redshift and magnitudes of the training and
photometric samples.  Note that the training sample
in this case is a faithful representation of the photometric
sample both in magnitude and in redshift.

\subsection{Photometric accuracy and precision metrics}
\label{sec:metrics}

The galaxies of the photometric sample have a set of
observables (the magnitudes measured from the five
standard SDSS filters \textit{ugriz}) along with the
spectroscopic redshift. Since the spectroscopic redshift 
is known in this case, we can evaluate the
accuracy and precision of the photometric redshift
estimation. Let $z_{\rm spec}$ and $z_{\rm phot}$ be the
spectroscopic and photometric redshifts, respectively, 
$\Delta z=z_{\rm phot}-z_{\rm spec}$ be the individual bias and $N$
be the number of photometric galaxies. Then, the metrics selected to
evaluate the photometric redshift estimation are as follows
(\cite{Sanchez2014}). \\
\\
(i) Bias

\begin{eqnarray}
\label{eq:bias}
\mu= \frac{1}{N} \sum_{i=1}^N (\Delta z_i )
\end{eqnarray}
\\
(ii) Dispersion

\begin{eqnarray}
\label{eq:sigma}
  \sigma = \sqrt{\frac{1}{N} \sum_{i=1}^N (\Delta z_i - bias)^2}
\end{eqnarray}
\\ 
(iii) Precision in 68-quantile: half the width of the distribution,
measured with respect to the median, in which 68\% of the
data points are enclosed. This is computed as

\begin{eqnarray}
\label{eq:sigma68}
\sigma_{\rm 68} = \frac{1}{2} (P_{\rm 84}-P_{\rm 16})
\end{eqnarray}
\\
where \\
$P_{\rm 16}$=16th percentile of the cumulative distribution \\
$P_{\rm 84}$=84th percentile of the cumulative distribution \\
\\
(iv) $f_{\rm 2\sigma}$: fraction of outliers above the $2\sigma$ level

\begin{eqnarray}
\label{eq:outliers2sigma}
f_{\rm 2\sigma}= \frac{1}{N} \sum_{i=1}^N W_i
\end{eqnarray}
\\
where

\begin{eqnarray}
W_i= \begin{cases} 1 & \mbox{if } {\left|\Delta z_i-bias\right|>2\sigma } \\ 0 & \mbox{if } {\left|\Delta z_i-bias\right|\leq 2\sigma }  \end{cases}
\end{eqnarray}
\\
(v) $f_{\rm 3\sigma}$: fraction of outliers above the $3\sigma$ level

\begin{eqnarray}
\label{eq:outliers3sigma}
f_{\rm 3\sigma}= \frac{1}{N} \sum_{i=1}^N W_i
\end{eqnarray}
\\
where,

\begin{eqnarray}
  W_i= \begin{cases} 1 & \mbox{if } {\left|\Delta z_i-bias\right|>3\sigma } \\ 0 & \mbox{if } {\left|\Delta z_i-bias\right|\leq 3\sigma }  \end{cases}
\end{eqnarray}
\\ 
(vi) $N_{\rm Poisson}$: a metric to quantify how close the
distribution of photometric redshifts $N(z_{\rm phot})$ is to the
distribution of spectroscopic redshifts $N(z_{\rm spec})$.

Let $N_i(z_{\rm phot})$ and $N_i(z_{\rm spec})$ be the relative number of photometric and spectroscopic galaxies, respectively, with redshifts in the $ith$ bin. Then
for each photometric redshift bin $i$ of width 0.05, we compute the difference of $N_i(z_{\rm phot})$ and $N_i(z_{\rm spec})$, normalized by the Poisson fluctuations on $N_i(z_{\rm spec})$.

\begin{eqnarray}
\label{eq:Npoisson}
  N_{\rm poisson}= \sqrt{\frac{1}{nbins}\sum_{i=0}^{nbins}\frac{(N_i(z_{\rm phot})-N_i(z_{\rm spec}))^2}{N_i(z_{\rm spec})}}
\end{eqnarray} 
\\
(vii) Kolmogorov-Smirnov test (KS): this statistic quantifies whether the two redshift 
distributions $N(z_{\rm phot})$ and $N(z_{\rm spec})$ are compatible with having been drawn from the same parent 
distribution.
\begin{eqnarray}
\label{eq:KS}
  KS=max(|F_{\rm phot}(z)-F_{\rm spec}(z)|)
\end{eqnarray} 
where $F_{\rm phot}(z)$ and $F_{\rm spec}(z)$ are the empirical cumulative distributions functions 
of $z_{\rm phot}$ and $z_{\rm spec}$, respectively.

\begin{figure*}
\centering
  \leavevmode
  \begin{tabular}{cc}
  \includegraphics[width=0.48\textwidth]{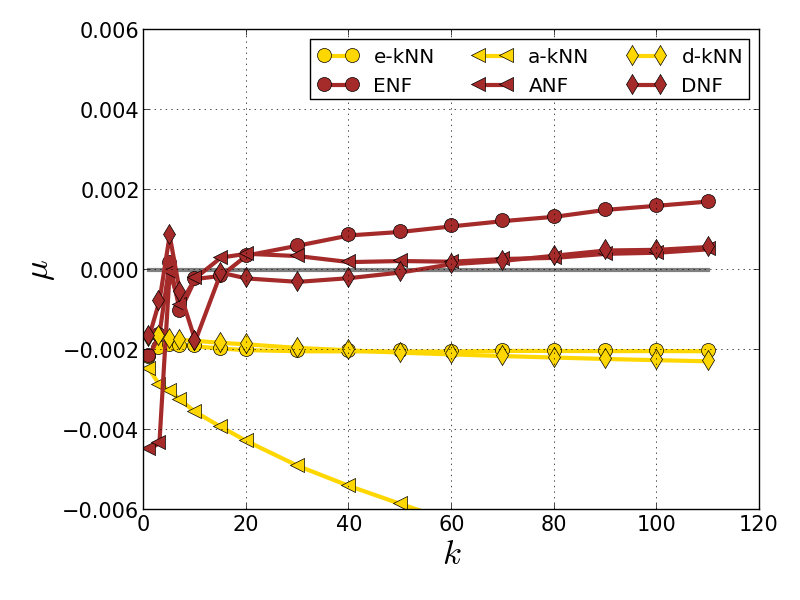} &
  \includegraphics[width=0.48\textwidth]{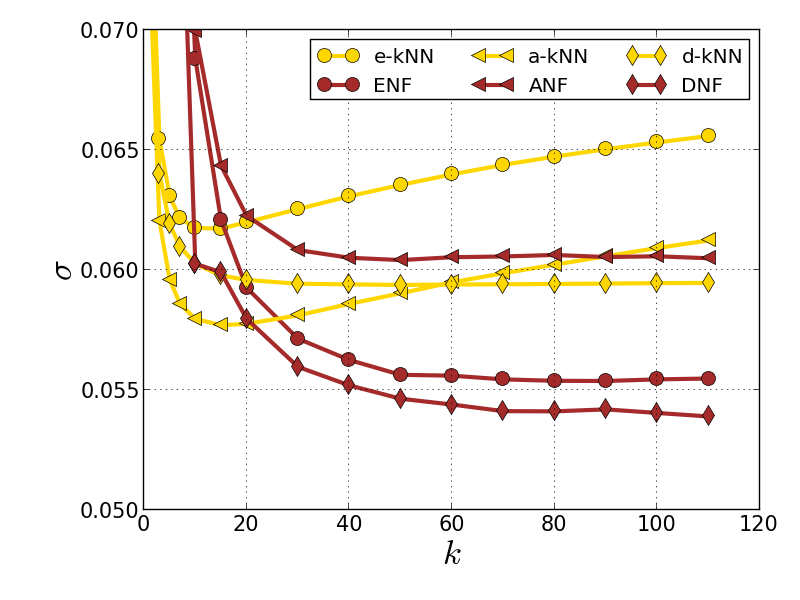} \\
  \end{tabular}
 \caption{Photo-z bias and dispersion metrics as a function of the number of neighbours for 
Euclidean (dot), angular (triangle) and directional (diamond) neighbourhoods. Light lines correspond to kNN algorithm 
explained in Section~\ref{sec:nncomparison} and dark lines to NF described in Section~\ref{sec:dnf}.
 }
  \label{fig:nneighbours}
\end{figure*}

\section{Nearest neighbour comparison}
\label{sec:nncomparison}

In this section, we use the kNN algorithm to compare the three
neighbourhoods described in Section~\ref{sec:nneighbourhood}: Euclidean, angular and directional. The
comparison is performed using the photometric galaxy
testbench defined above.

For kNN photometric redshift estimation we select the $k$ nearest neighbours in
magnitude space.  The photo-z is then computed
as the mean of the redshift of these $k$ nearest neighbours,
pondered inversely to the corresponding distance metric of each neighbour:

\begin{eqnarray}
z_{\rm phot}=\alpha \sum_{i=1}^{k} \frac{z_i}{d_i} ,
\label{eq:meaneq}
\end{eqnarray} 

\noindent
where $\alpha$ is the normalization constant given by

\begin{eqnarray}
\alpha = \frac{1}{\sum_{i=1}^{k} 1/d_i}
\label{eq:eqweights}
\end{eqnarray} 

To determine the optimum number of neighbours ($k$) to average,
we studied the different metrics as a function of the
number of neighbours for a random subsample of 50,000
galaxies. The kNN graphs in Fig.~\ref{fig:nneighbours}
 show that the bias is stable for
e-kNN (light dots) and d-kNN (light diamonds), whereas for a-kNN (light triangles) it grows with the number of
neighbours. Regarding the dispersion, it is
low and stable beyond 20 neighbours for d-kNN, while for e-kNN  and a-kNN, minimums of around 15 and 20
neighbours, respectively, are found. From these plots, d-kNN
 appears to provide the best compromise between accuracy and precision
 for the kNN approach.

\begin{figure*}
\centering
  \leavevmode
 \begin{tabular}{cc}
  \includegraphics[width=0.48\textwidth]{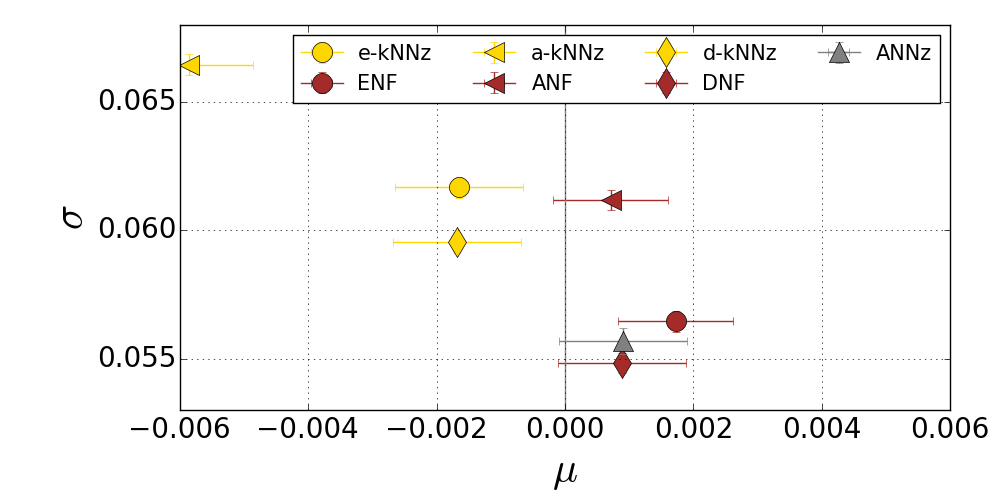} &
  \includegraphics[width=0.48\textwidth]{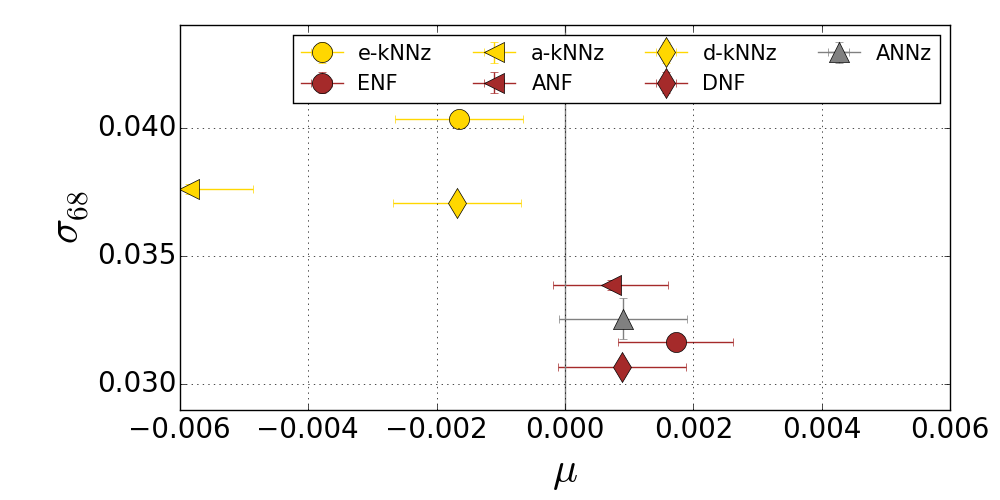} \\
  \includegraphics[width=0.48\textwidth]{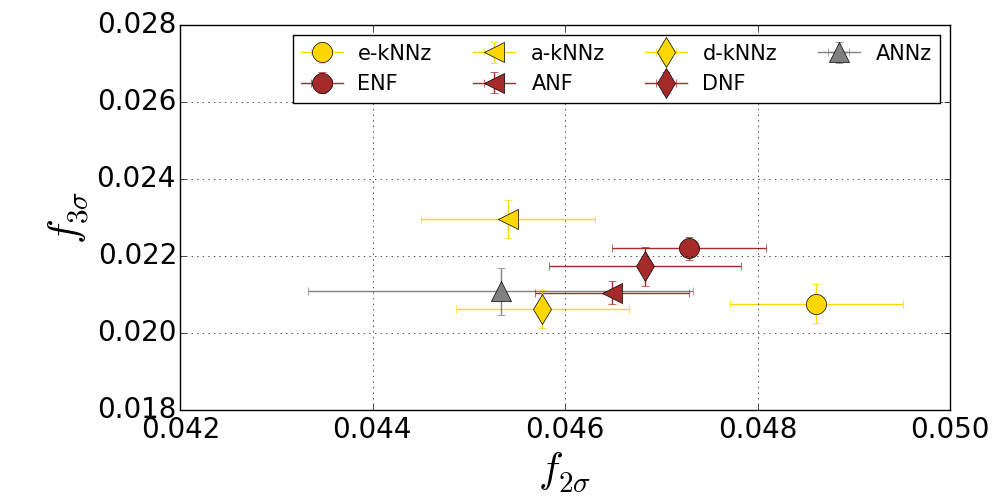} &
  \includegraphics[width=0.48\textwidth]{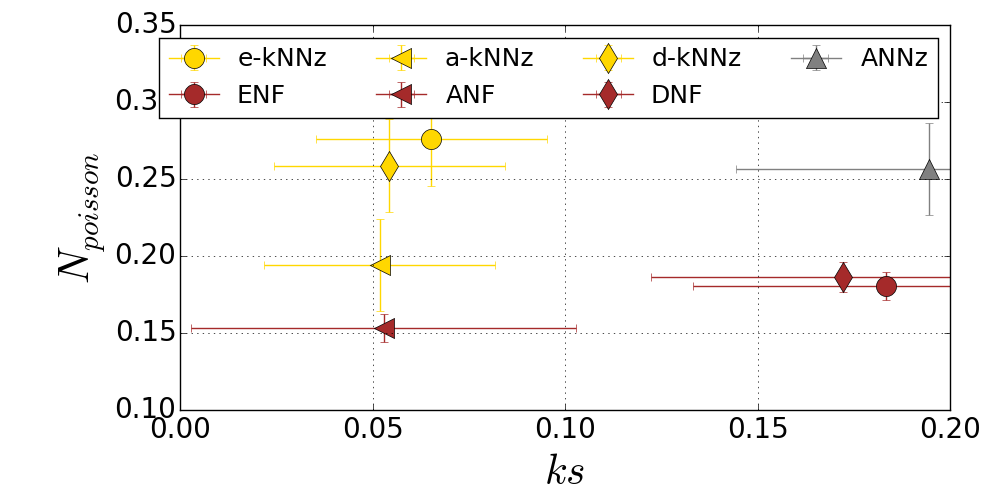} \\
 \end{tabular}
 \caption{Comparison of different combination of nearest neighbour strategies (k-NN and NF) 
and neighbourhoods (dot: Euclidean; triangle: angular; diamond: directional). Bias versus $\sigma$, bias versus $\sigma_{\rm 68}$, $f_{\rm 2\sigma}$ 
versus $f_{\rm 3\sigma}$ outliers, and $N_{\rm Poisson}$ versus KS are presented. The plots also show the results produced by the reference
code ANNz. DNF produces less bias and scatter than the other codes.}
  \label{fig:metrics1}
\end{figure*}

\section{Directional Neighbourhood Fitting}
\label{sec:dnf} 

In this section, we describe DNF, a new strategy that clearly
improves the kNN approaches. DNF computes the photo-z hyperplane
that best fits the directional neighbourhood of a photometric
galaxy in the training sample. This hyperplane serves as a
prediction function for the photo-z of the galaxy
considered. Note that since the fitting is local, a linear
adjustment (hyperplane) should sufficient to make a fine
prediction. Let $\textbf{\textit{m}}$ be a multi-magnitude vector in the
neighbourhood of the photometric galaxy $G$. Then, a generic
relationship can be established between redshifts ($z$) and
magnitudes ($\textbf{\textit{m}}$) in the neighbourhood of $G$ as shown in equation ~\ref{eq:photoz}:

\begin{eqnarray}
\label{eq:photoz}
z= \mathbf{\mathit{a}} \cdot \mathbf{\mathit{m}}
\end{eqnarray}

where $\textbf{\textit{a}}$ is a parameter vector. This can be computed as the best fit
between redshifts and magnitudes of the galaxies in the training
sample that are neighbours of $G$.  Once $\textbf{\textit{a}}$ is
computed, equation ~\ref{eq:photoz} is applied for the
photo-z prediction of the galaxy $G$. In addition, the residual of the fit 
can be considered the photo-z estimated error. The NF
method is also applied to the Euclidean and angular
neighbourhoods for comparison. We have denoted these methods
as Euclidean neighbourhood fitting (\textbf{ENF}) and
angular neighbourhood fitting (\textbf{ANF}),
respectively. Regarding the number of neighbours ($k$) used
in the fit, a compromise must be found between higher
statistical power (larger $k$) and locality (smaller
$k$). Fig.~\ref{fig:nneighbours} shows the dependency of $\mu$ 
and $\sigma$ on $k$ for the different neighbourhood fitting
methods. From the figure, it is noteworthy, that the NF 
approaches admit a large number of neighbours, and that both accuracy
and precision are generally better than obtained with kNN. The difference may be due to a non-optimal 
weighting scheme for kNN based on the inverse of distance ($1/d$).
In addition, DNF (dark diamonds) behaves better than ENF (dark dots) for any value of $k$.
 Thus, ENF bias grows faster with $k$ than it does in DNF. 
ENF also shows larger scatter ($\sigma$) in the computed photo-zs than DNF .

\subsection{Neighbourhood fitting vs. k-nearest neighbour}

In this subsection, we compare the kNN and NF approaches for the
entire photometric sample ($\sim 1$ million galaxies). The value of
$k$ is fixed at an appropriate value for each method selected
in the view of Fig.~\ref{fig:nneighbours} (25 for kNN and 60
for NF). As a reference, we have also included the results produced by 
the well-known ANNz algorithm (\cite{2004PASP..116..345C}). 

Fig.~\ref{fig:metrics1} shows the result of this
comparison. Note that $\mu$, $\sigma$ and $\sigma_{\rm 68}$
perform better for NF than they do for kNN, and at a level close to the reference ANNz. Among the NF
approaches, DNF stands out as superior to the other two neighbourhoods
for the metrics considered.
 
\begin{figure*}
  \centering
  \leavevmode
  \begin{tabular}{cc}
    \includegraphics[width=0.5\textwidth]{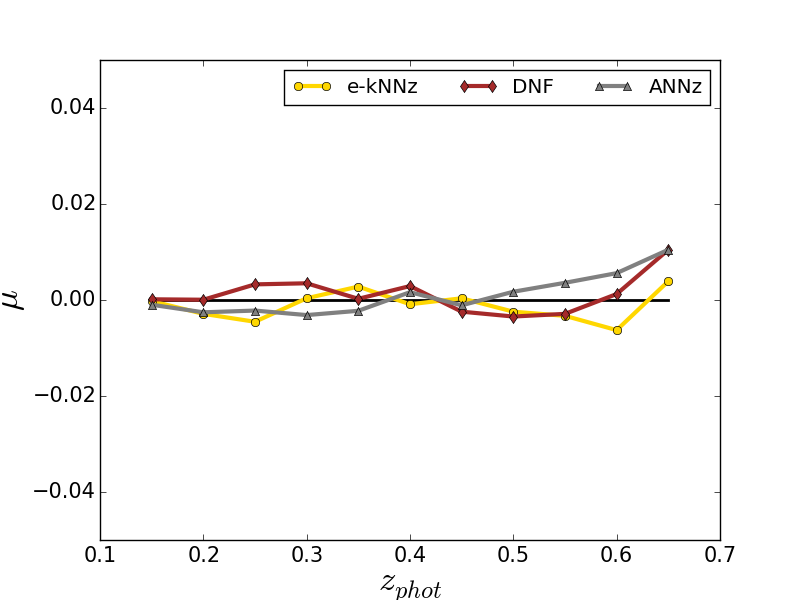} &
    \includegraphics[width=0.5\textwidth]{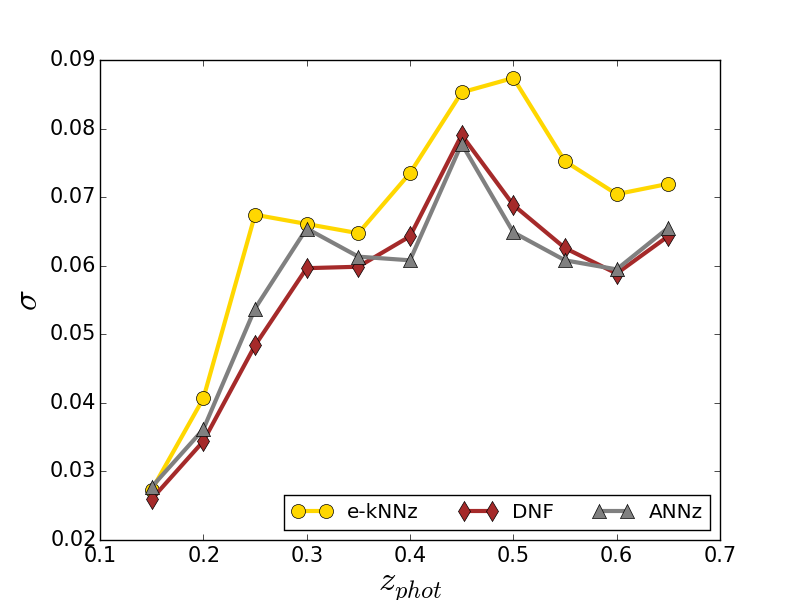} \\
    \includegraphics[width=0.5\textwidth]{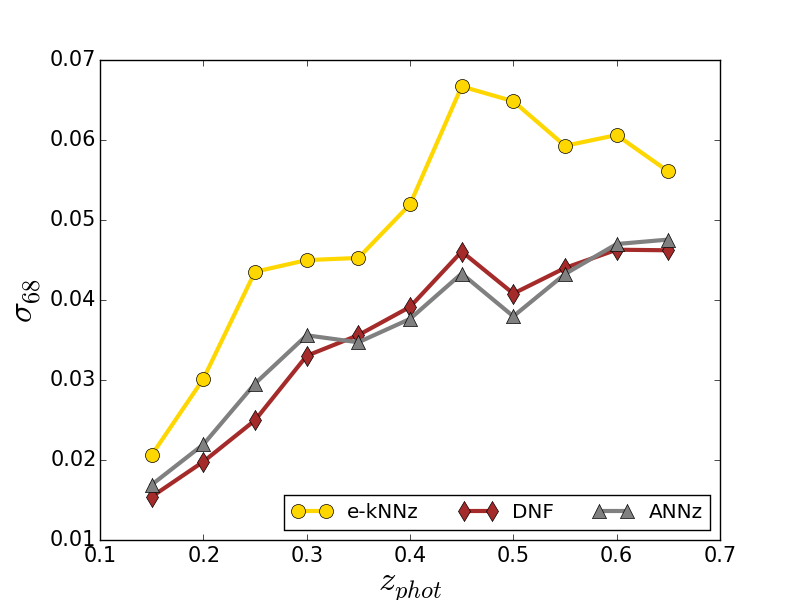} &
    \includegraphics[width=0.5\textwidth]{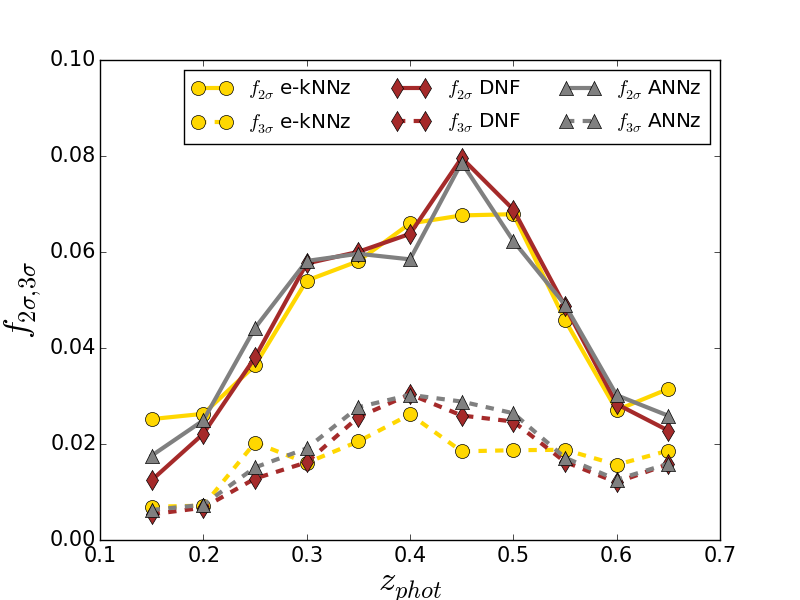} \\
  \end{tabular}
  \caption{Dependence on the redshift of the metrics for the photometric sample for the codes kNN, DNF and ANNz. Note 
  the improvement of DNF with respect to kNN while a similar performance to ANNz is achieved.}
  \label{fig:sbvsz}
\end{figure*}

Regarding the dependency on the redshift, we focused the
study on only three methods: e-kNN (i.e. the common kNN approach), the newly defined
DNF method, and the reference ANNz. Fig.~\ref{fig:sbvsz} shows the comparison in bins of
photo-z. DNF behaves better than kNN and similar to ANNz.
 
\begin{figure*} 
  \centering
  \leavevmode
  \begin{tabular}{ccc}
    \includegraphics[width=0.33\textwidth]{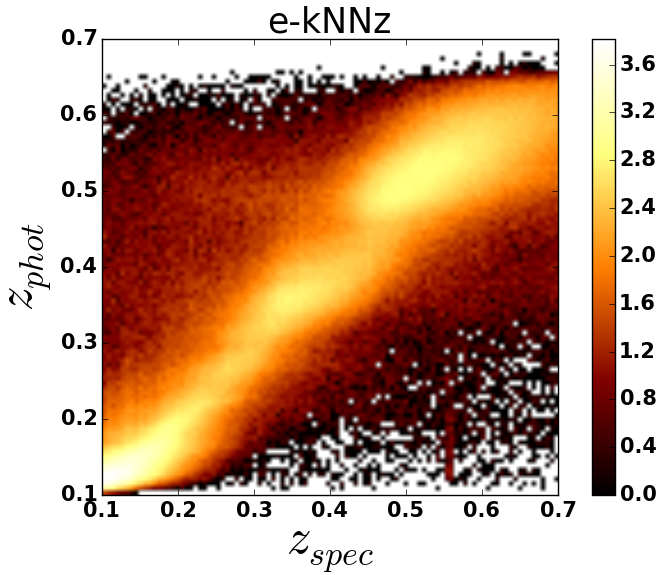} &
    \includegraphics[width=0.33\textwidth]{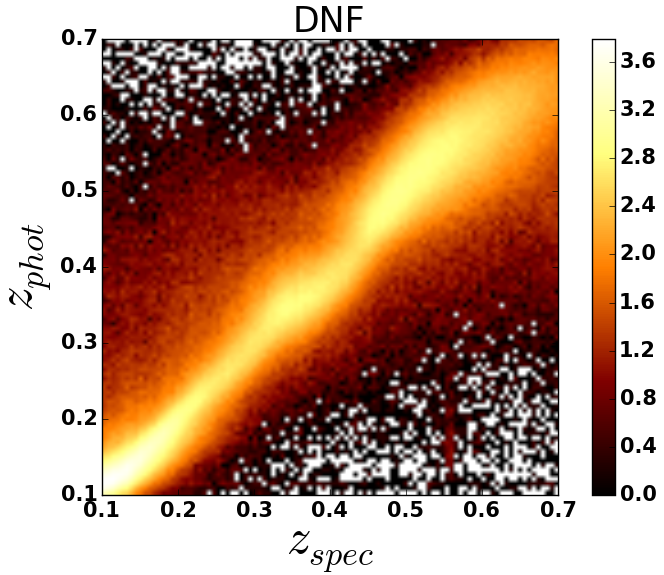} &
    \includegraphics[width=0.33\textwidth]{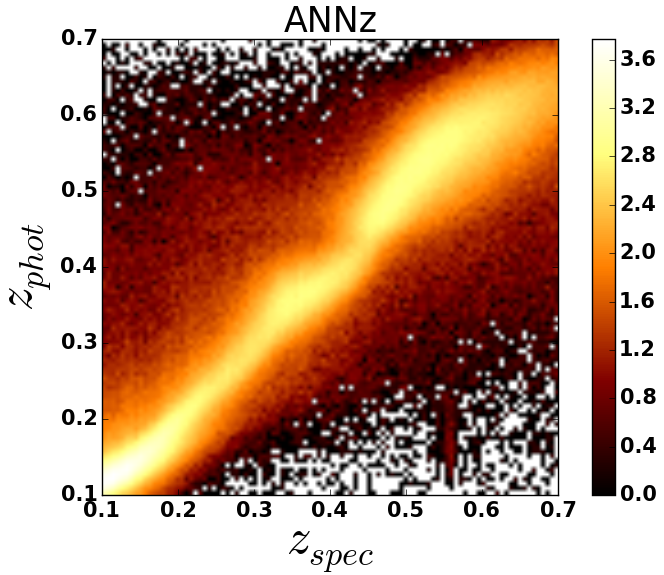}  \\
  \end{tabular}
  \captionsetup{labelfont=bf} \captionof{figure} {Scatter plot ($z_{\rm photo}$ versus $z_{\rm spec}$) for the photometric sample produced by 
 e-kNN (left), DNF (centre) and ANNz (right). Note that DNF clearly improves kNN dispersion and behaves similarly to ANNz.}
  \label{fig:scatter}
\end{figure*}

In Fig. ~\ref{fig:scatter} we can see the scatter plot
showing the spectroscopic redshifts ($z_{\rm spec}$) versus the
estimated photometric redshifts ($z_{\rm phot}$) for the three
methods. The DNF approach clearly improves the dispersion
seen with kNN and behaves similarly to ANNz for this sample.

With respect to the photo-zs on the edges of the training
sample some caveats should be noted. Usually, the redshift
limits of the training sample are less populated by
galaxies. Finding a fixed number of neighbours in these
regions may require moving further away and expanding into
regions where the fit is no longer reliable. We have assumed
that photo-zs that go beyond the redshift limits of the
training sample are the result of failed non-local
fitting. Accordingly, a correction has been applied using the
redshift of the nearest neighbour. Another way to circumvent non-local fitting is by limiting the neighbour
selection to within a hyper-volume of small radius
(or directional radius for directional
  neighbourhoods). In this approach, the number of neighbours
($k$) depends on the galaxy density of the multi-magnitude
space region, and thus, the proper algorithm should be
applied. According to Fig.~\ref{fig:nneighbours}, DNF is
appropriate for moderate and large values of $k$ but the
fitting process may fail for low values (e.g. $k<20$). In
these cases d-kNN seems to be a safer approach and a
hybrid method can be implemented. Finally, it should be noted that a
weighting technique, as detailed in
\cite{2008MNRAS.390..118L}, can also be implemented so that
the training sample resembles the photometric sample. This can be done in
order to achieve higher representativeness when computing photo-zs.

\subsection{Photo-z probability density functions}
\label{sec:photozpdfs}
In addition to a single $z_{\rm phot}$ value, DNF can produce
PDFs which better reflect
the uncertainty associated with the photo-z prediction. Let
$s_i$ be the residual of the $i$-th neighbour generated in
the fit to the hyperplane (equation ~\ref{eq:photozz}).  

\begin{eqnarray}
\label{eq:photozz}
s_i= z^i_{\rm spec} - \mathbf{\mathit{a}} \cdot \mathbf{\mathit{m_i}}
\end{eqnarray}

For each neighbour a PDF sample $z_i$ is
generated by summing the residuals $s_i$ to $z_{\rm phot}$ according
to equation ~\ref{eq:zdist}.  

\begin{eqnarray}
\label{eq:zdist}
z_i= z_{\rm phot}+s_i
\end{eqnarray}

\begin{figure*}
  \centering
  \leavevmode
  \begin{tabular}{ccc}
    \includegraphics[width=0.3\textwidth]{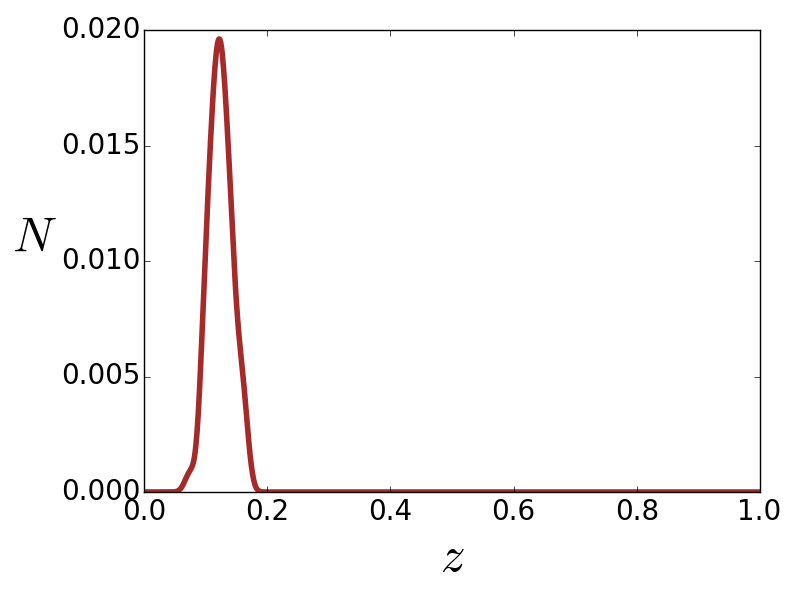} &
    \includegraphics[width=0.3\textwidth]{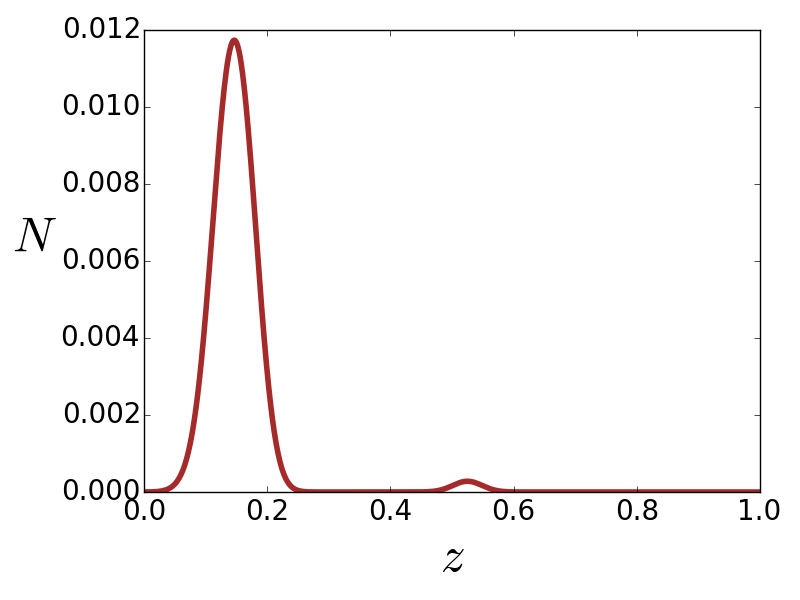} &
    \includegraphics[width=0.3\textwidth]{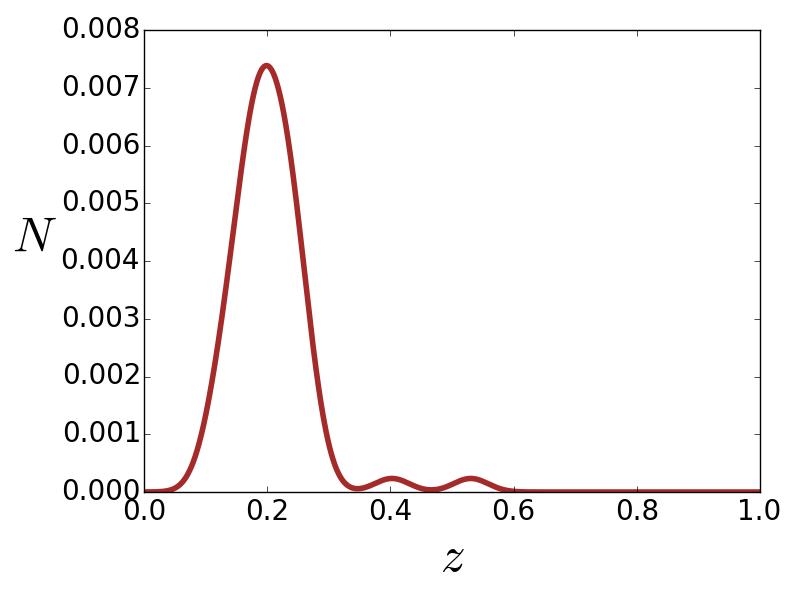} \\
    \includegraphics[width=0.3\textwidth]{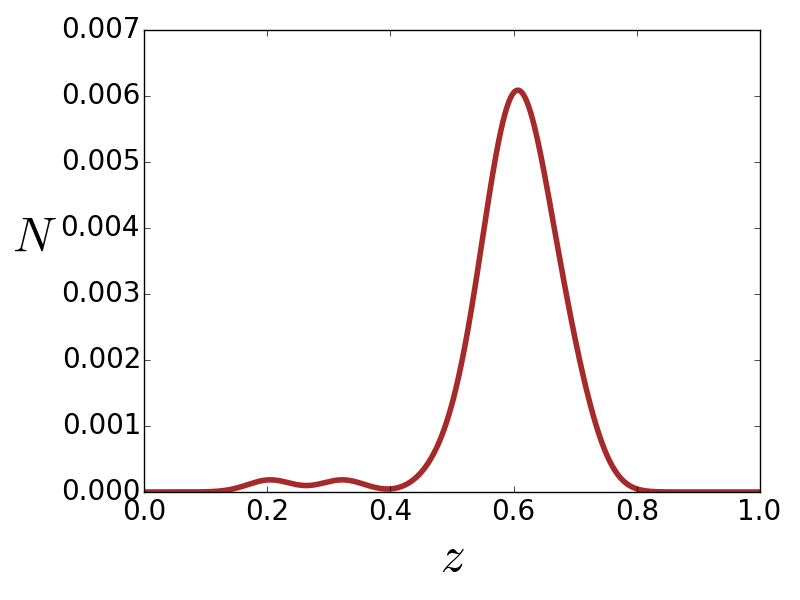} &
    \includegraphics[width=0.3\textwidth]{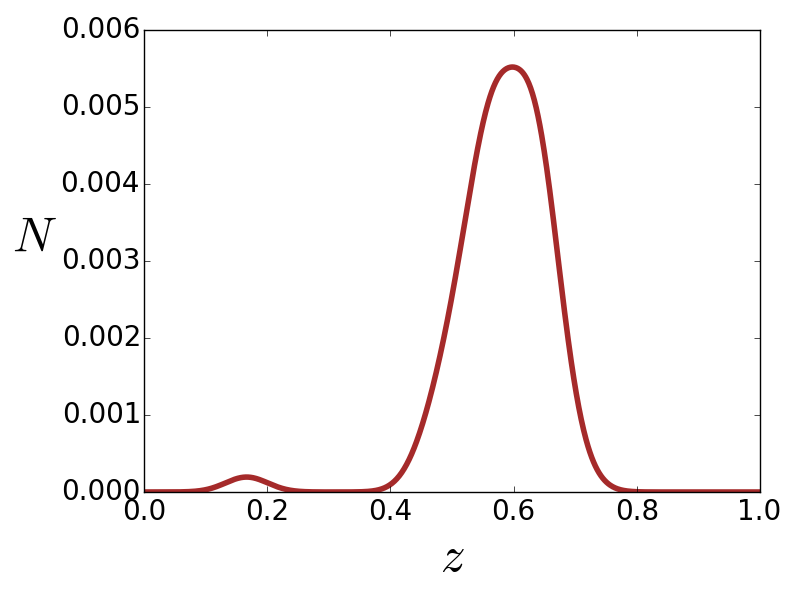} &
    \includegraphics[width=0.3\textwidth]{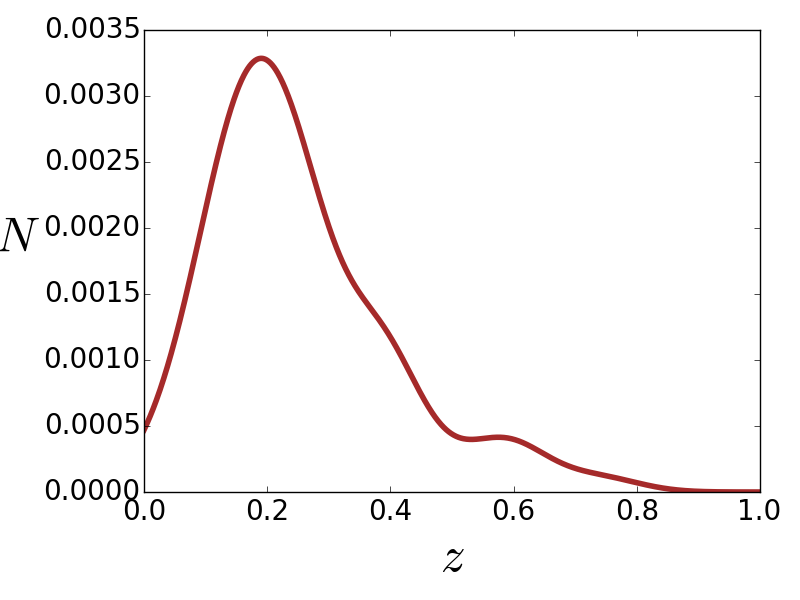} \\
  \end{tabular}
  \caption{Examples of individual photo-z PDF generated by DNF code. Note that the PDF enables the algorithm to capture the uncertainty and degeneracies.}
  \label{fig:galaxypdf}
\end{figure*}

The PDF is then obtained by
histogramming these samples in redshift bins.
Fig.~\ref{fig:galaxypdf} provides some examples of individual photo-z PDFs obtained by Kernel Density Estimation (KDE)
applied to the PDF samples. Note that non-Gaussian uncertainty distributions and degeneracies are reflected by the PDFs
computed by DNF.

\begin{figure*}
    \centering
    \leavevmode
      \includegraphics[width=1.0\textwidth]{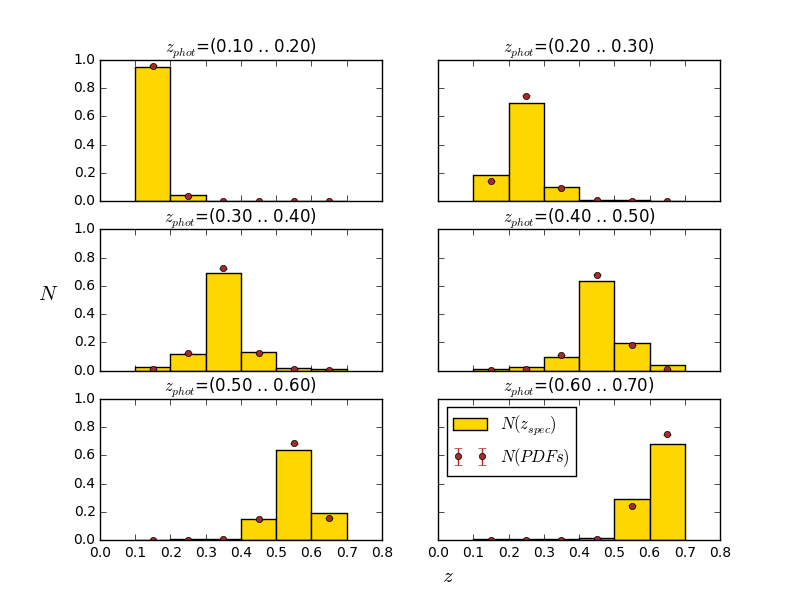}
      \caption{$N(z_{\rm spec})$ and $N(PDFs)$ both in $z_{\rm phot}$ bins (i.e. galaxies are first classified into 
	bins according to their photo-z and then a comparison between the $z_{\rm spec}$ distribution and the PDF stacking is performed for each bin).}
      \label{fig:Nza}
 \end{figure*}

Fig.~\ref{fig:Nza} shows how the photo-z PDFs also allow us to estimate
the uncertainty of the photometric-sample distribution in $z_{\rm phot}$ bins.
The photometric sample distribution (bars in the graph) in each $z_{\rm phot}$ bin is constructed by: (a) classifying 
the galaxies in bins according to its $z_{\rm phot}$, and (b) histogramming the $z_{\rm spec}$ of the galaxies
classified in the target $z_{\rm phot}$ bin. Estimation of the photometric sample distribution, on the other hand, is done
 by stacking the photo-z PDFs ($N(PDFs)$) classified in the target bin (dots in the graph).

\begin{figure*}
      \centering
      \leavevmode
      \includegraphics[width=0.7\textwidth]{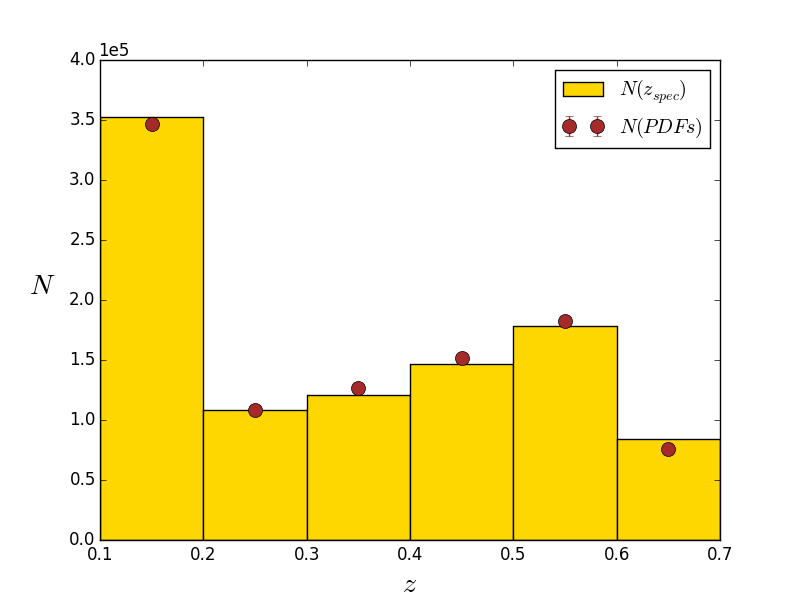}
      \caption{$N(z_{\rm spec})$ versus $N(PDFs)$ for the photometric sample. Note that the measured distribution is well estimated by stacking photo-z PDFs. }
      \label{fig:Nzb}
\end{figure*}

Fig.~\ref{fig:Nzb} shows $N(z_{\rm spec})$ (the global photometric-sample distribution) and $N(PDFs)$.
 Note that the shape of the global redshift distribution
 is well reproduced by stacking the PDFs.

\subsection{Photo-z error estimation}
\label{sec:zerrestimation}
In addition to photo-z computation, an important issue is the associated
error estimation. In kNNz approaches, the standard
deviation for the redshift of the \textit{k} nearest neighbours
in the training sample is considered the photo-z error estimation for the
galaxy. In NF methods, as we have already established,
 the residuals of the fit provide a good estimation of the photo-z bias.

\begin{figure*} 
  \centering
  \leavevmode
  \begin{tabular}{cc}
    \includegraphics[width=0.5\textwidth]{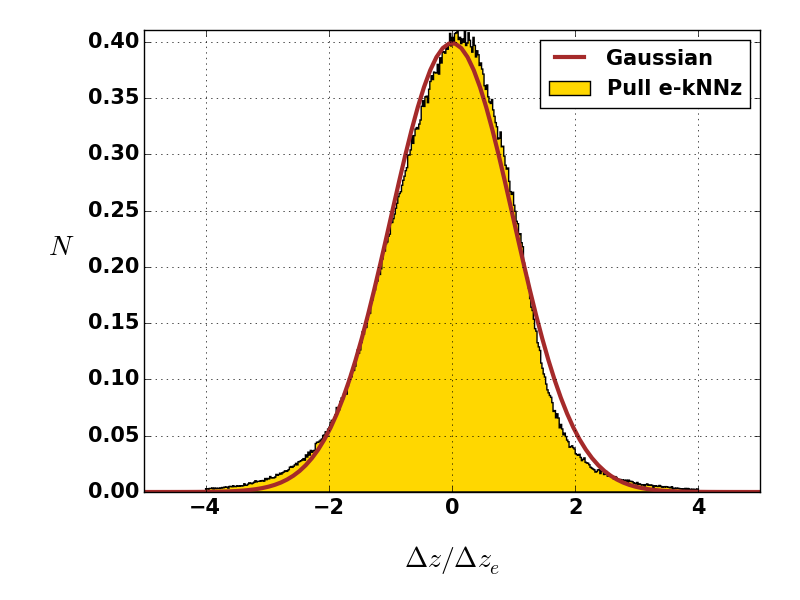} &
    \includegraphics[width=0.5\textwidth]{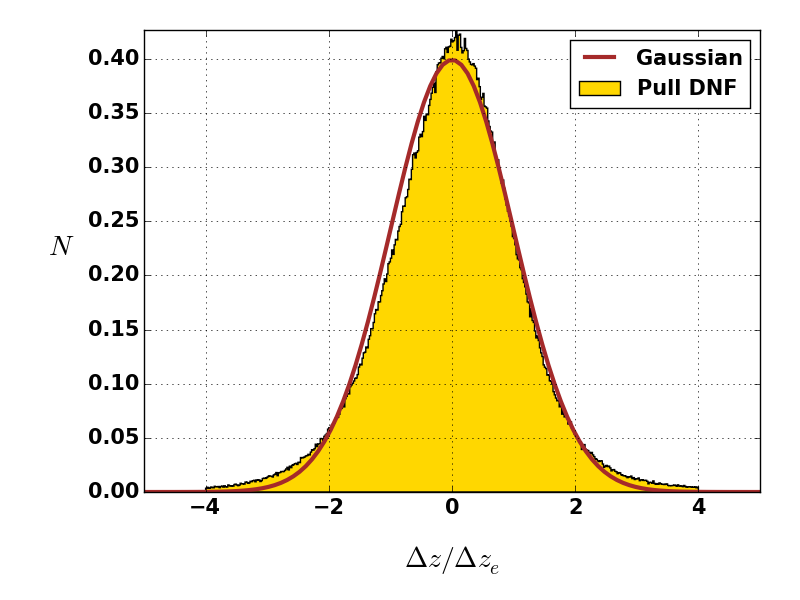} \\
  \end{tabular}
  \captionsetup{labelfont=bf} \captionof{figure}{Pull: $\Delta z/\Delta z_e$ distribution for e-kNN (left) and DNF (right). 
Also drawn is the ideal pull distribution, which is a Gaussian, with $\mu=0$ and $\sigma=1$.}
  \label{fig:pull}
\end{figure*}

Fig. \ref{fig:pull} shows the distribution for the e-kNNz and DNF codes
of $\Delta z/\Delta z_e$ (pull distribution), where $\Delta z$ is the
individual bias and $\Delta z_e$ is the photo-z
error estimation. Note that the ideal pull distribution, , which is also drawn, is
Gaussian, with $\mu=0$ and $\sigma=1$.

\begin{figure*} 
  \centering
  \leavevmode
  \begin{tabular}{ccc}
    \includegraphics[width=0.3\textwidth]{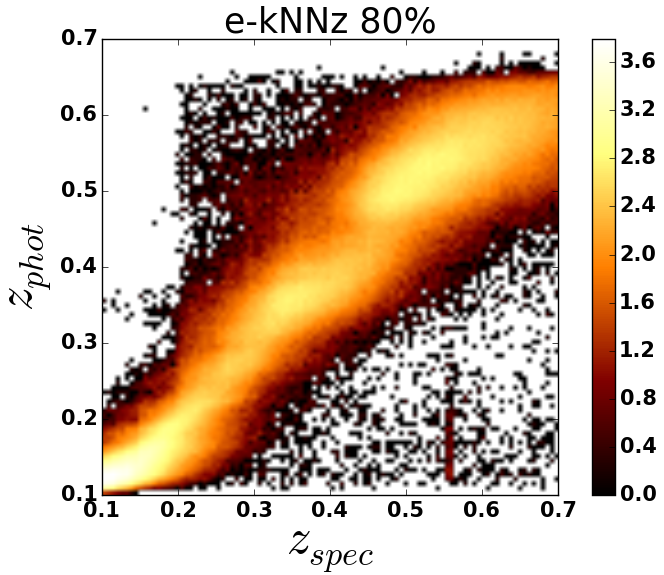} &
    \includegraphics[width=0.3\textwidth]{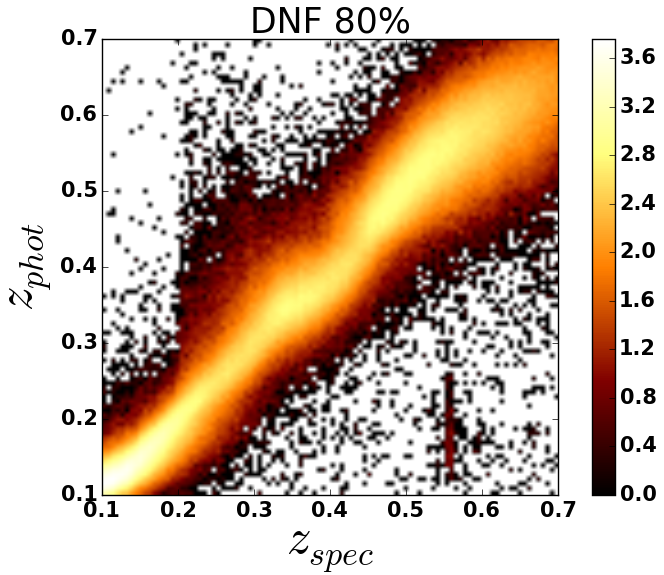} &
    \includegraphics[width=0.3\textwidth]{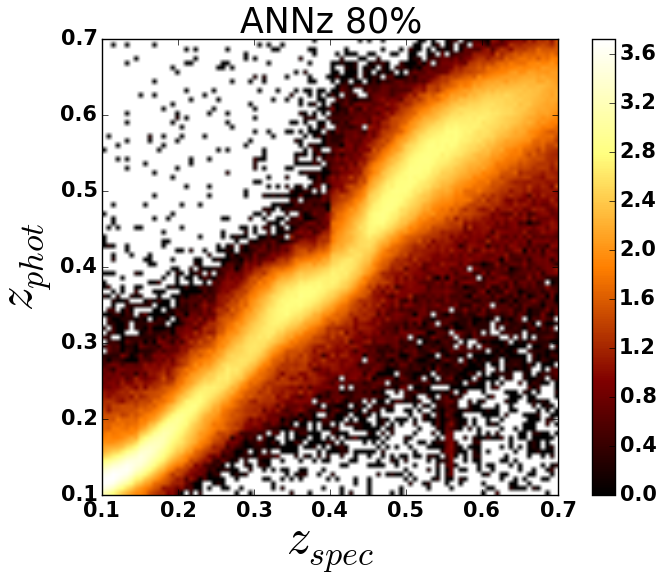}\\
  \end{tabular}
  \captionsetup{labelfont=bf} \captionof{figure} {Scatter plot ($z_{\rm photo}$ versus $z_{\rm spec}$) for the photometric sample 
with a quality cut of the $80\%$ of the galaxies with the best photo-z error estimation, applied in bins of $\Delta z=0.05$, and provided by the codes e-kNN (left),
 DNF (centre) and ANNz (right).}
  \label{fig:scatterQC80}
\end{figure*}

Another way to visualize the suitability of the photo-z error estimation is by performing a quality cut based on this parameter. Fig. \ref{fig:scatterQC80}  
shows the scatter plot for kNN and DNF, created by selecting the 80\% of galaxies per redshift bin that have a smaller photo-z 
estimated error. Note the good behaviour of DNF, as many of the outliers have been removed.

\section{Application of DNF to other survey samples} 
\label{sec:othersurveys}
So far, we have applied DNF to the SDSS DR10 which provides  a large number of galaxy spectroscopic redshifts to compare it with.
Nevertheless, SDSS DR10 has a very limited redshift range. In this section we extend the comparison to higher redshift. 
First, we apply DNF to the VVDS sample up to $z=1$. Next, we apply DNF to the public PHAT dataset.

\subsection{DNF on VVDS}
\label{sec:dnfonvvds}

\begin{figure*}
\centering
  \leavevmode
 \begin{tabular}{cc}
  \includegraphics[width=0.48\textwidth]{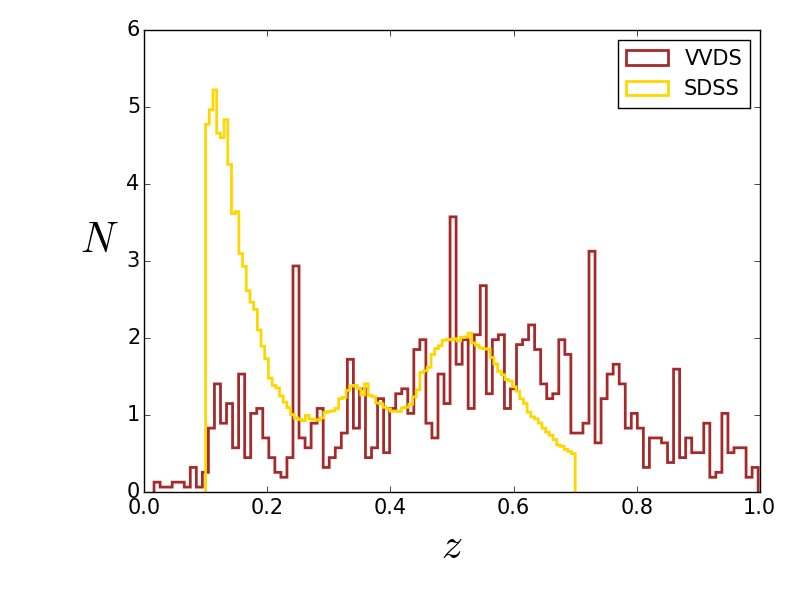} &
  \includegraphics[width=0.48\textwidth]{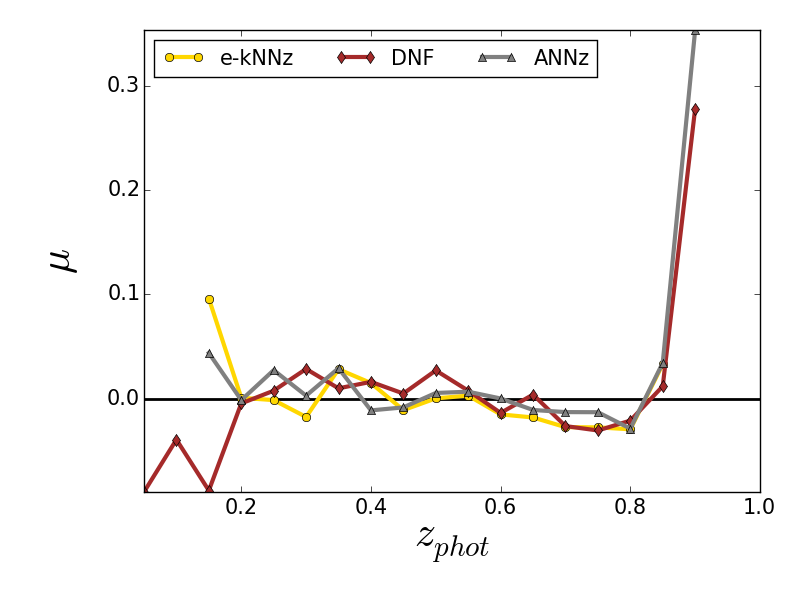} \\
  \includegraphics[width=0.48\textwidth]{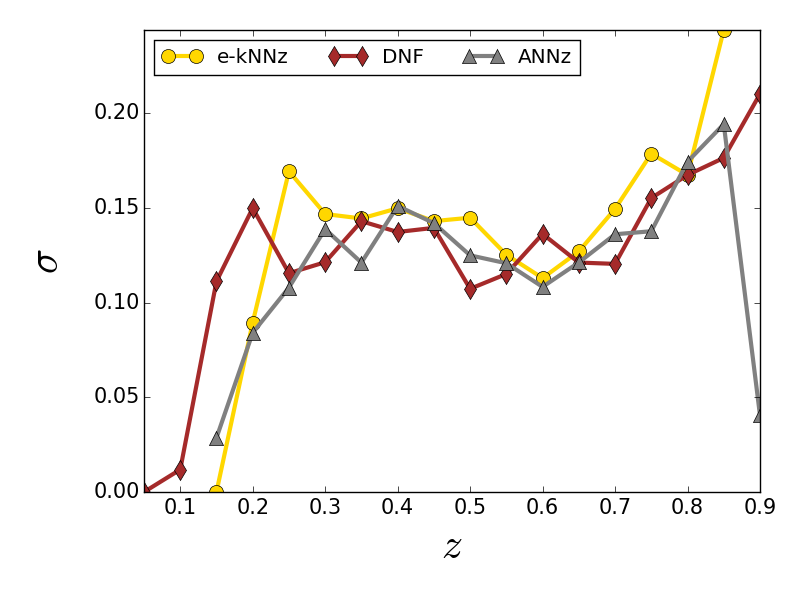} &
  \includegraphics[width=0.48\textwidth]{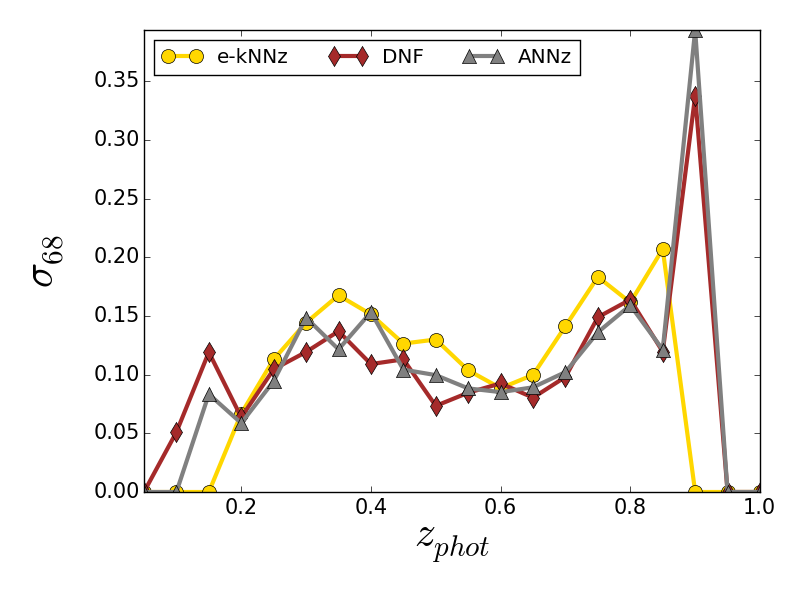} \\
 \end{tabular}
\caption{VVDS redshift distribution $N(z_{\rm spec})$ of the photometric sample compared to the SDSS sample (upper left). The rest of 
the figures show the dependence of the metrics on $z$ for the codes kNN, DNF and ANNz. 
Note how DNF behaves similar to ANNz in the central part of the redshift range. The strange behaviour of the plots at the redshift
 extremes is explained by the shortage of training galaxies.}
  \label{fig:vvdsdistribution}
\end{figure*}

\begin{figure*}
\centering
  \leavevmode
 \begin{tabular}{cc}
  \includegraphics[width=0.48\textwidth]{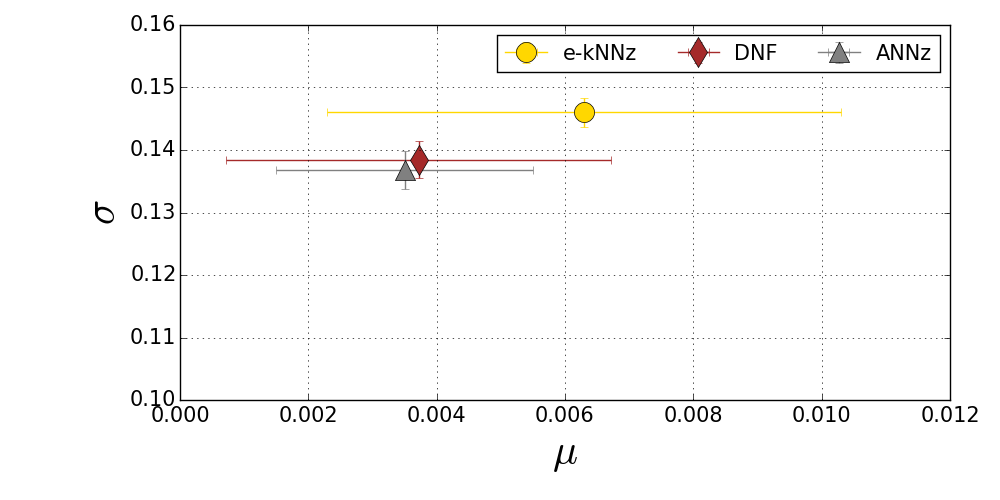} &
  \includegraphics[width=0.48\textwidth]{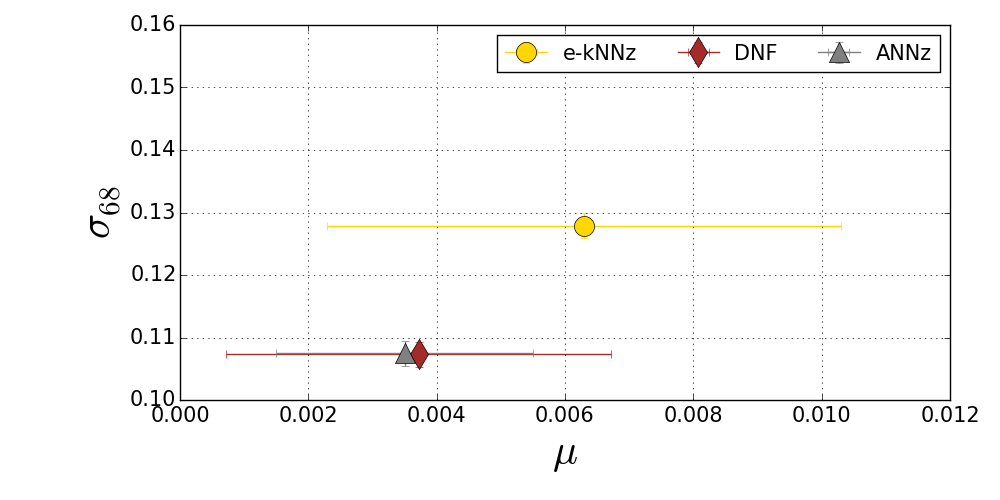} \\
  \includegraphics[width=0.48\textwidth]{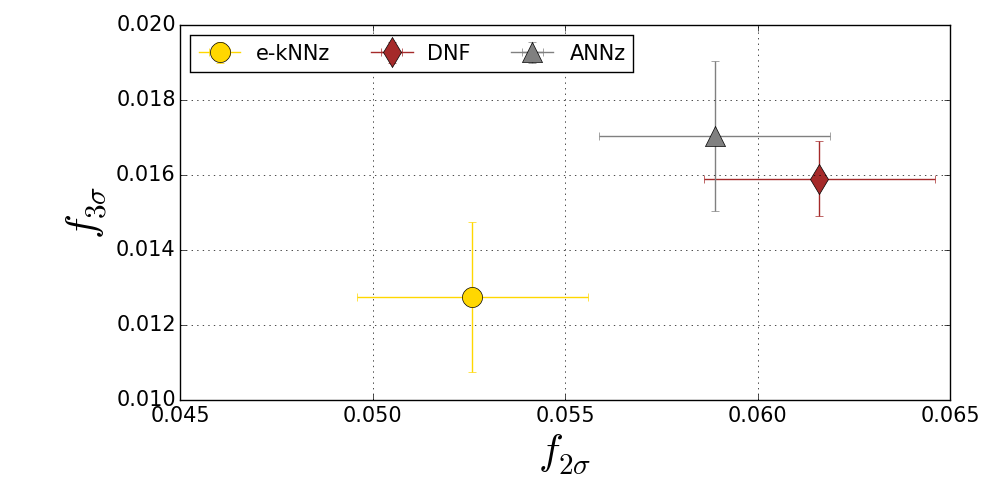} &
  \includegraphics[width=0.48\textwidth]{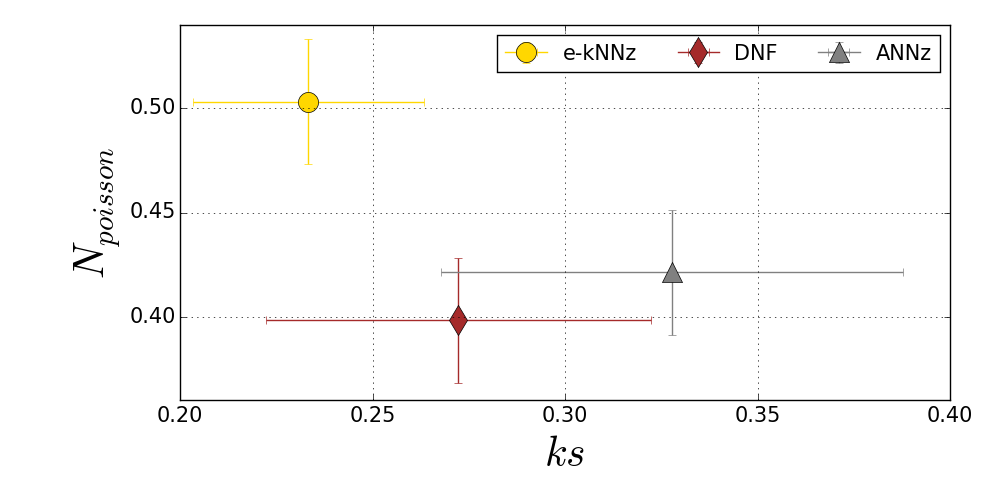} \\
 \end{tabular}
  \caption{Comparison among kNN, DNF and ANNz for the VVDS sample up to $z=1$. Bias versus $\sigma$, bias versus $\sigma_{\rm 68}$, $f_{\rm 2\sigma}$ 
versus $f_{\rm 3\sigma}$ outliers and $N_{\rm Poisson}$ versus KS are represented.} 
  \label{fig:vvdsplots}
\end{figure*}

The VVDS is a comprehensive deep galaxy spectroscopic redshift survey conducted by the Visible Imaging Multi-Object Spectrograph (VIMOS)
 collaboration with the VIMOS multi-slit spectrograph at the ESO-Very Large Telescope (VLT; \cite{VVDS2008}). To evaluate the performance of DNF, we selected
3598 object from the VVDS-F1400+05 dataset. The selected objects are those in the redshift range (0, 1) and with 
valid magnitude values in the bands. We split the set into training and validation samples with 2,000 and 1,598 objects 
respectively. Fig.~\ref{fig:vvdsdistribution} shows the distribution of redshifts compared to the SDSS sample, and the dependence of the
metrics on the redshifts for the codes kNNz, DNF, and ANNz. Note how DNF behaves similarly to ANNz in the central part of the redshift 
range. The poor behaviour of the codes at the redshift extremes is explained by the shortage of training 
galaxies. In another view, Fig.~\ref{fig:vvdsplots} provides a summary comparison of the methods. Unlike with kNN, note how DNF 
achieves accuracy and precision close to ANNz for this set.

\subsection{DNF on PHAT} 
\label{sec:dnfonphat}

The PHAT program is an international initiative whose goal is to test and compare different methods of photo-z 
estimation. Two different test environments are set up: PHAT0, based on data produced by simulations; and PHAT1, based on data from the 
Great Observatories Origins Deep Survey (GOODS), including 18-band photometry and ~2,000 spectroscopic redshifts. We applied DNF to the PHAT data and evaluated its performance.
 To provide a fair comparison, we focused only on the training-based methods described in the PHAT paper (note the small size of 
the training set of ~500 objects). The methods used are the following: AN-e (ANNz, artificial neural network), DT-e (BDT, 
boosted decision trees), EC-e (empirical $\chi2$), PN-e (nearest neighbour fit), PO-e (polynomial fit), RT-e (regression trees),
 SN-e (neural network). Table ~\ref{table:phat} (left) duplicates the PHAT0 comparison (table 3 of the PHAT paper) along with DNF results, 
where bias and scatter are the mean and the RMS of the quantity $\Delta z = z_{\rm model}-z_{\rm phot}$ (after rejection of outliers) 
respectively, and the outlier are defined as objects with $\mid\Delta z\mid =\mid z_{\rm model}-z_{\rm phot}\mid>0.1$.
 Table ~\ref{table:phat} (right) duplicates the PHAT1 comparison (table 5, columns 5-7 of the PHAT paper) along with DNF results. In this case bias and scatter are the mean and rms of the quantity
$\Delta z' = (z_{\rm spec}-z_{\rm phot})/(1+z_{\rm spec})$ respectively and the outliers are defined as objects with $\mid\Delta z'\mid>0.15$. 
Note that DNF are among the best scored training approaches in both the PHAT0 and PHAT1 tests. 

\begin{table*}
\caption{DNF applied to PHAT data: PHAT0 (left) corresponds to a simulation data. PHAT1 (right) corresponds to GOODS field.}
\begin{center}
 \begin{tabular}{||c c c c || c c c c||}
 \hline
 \multicolumn{4}{||c||}{PHAT0} & \multicolumn{4}{c||}{PHAT1} \\
 \hline\hline
 Code & Bias & Scatter & Outlier (\%) & Code & Bias & Scatter & Outlier (\%) \\ [0.5ex] 
 \hline\hline
 AN-e &  0.000 & 0.011 & 0.018  & AN-e &  -0.006 & 0.078 & 38.5 \\ 
 \hline
 DT-e & -0.004 & 0.019 & 0.389 &  EC-e &  -0.002 & 0.066 & 16.7 \\
 \hline
 PN-e &  0.000 & 0.017 & 0.053 &  PO-e &  -0.007 & 0.051 & 13.0 \\
 \hline
 PO-e &  0.001 & 0.019 & 1.669 &  RT-e &  -0.008 & 0.067 & 24.2 \\
 \hline
 RT-e &  0.000 & 0.013 & 0.010 &  DNF-e &  0.004 & 0.063 & 21.0\\ 
 \hline
 SN-e & -0.005 & 0.049 & 18.202 &     &        &       &      \\
 \hline
 DNF-e & 0.000 & 0.011 & 0.026 &     &        &       &    \\
 \hline
\end{tabular}
\end{center}
\label{table:phat}
\end{table*}

\section{Conclusions}
\label{sec:conclusions} 

The precision of many cosmological probes relies on the
quality of the photometric redshift technique. Thus, great
effort has been put into developing new methods and improving established
techniques for both template and training based approaches.

In this paper, we have explored the application of nearest neighbours algorithm
to photometric redshift estimation. Two new
neighbourhoods (angular and directional) were proposed.
 The newly defined neighbourhoods, along with
the most commonly used, Euclidean neighbourhood, were combined with two
nearest neighbour approaches (kNN and NF) to form six methods. In 
addition the well-known ANNz was used as a reference.

The comparison of methods reveals the value of the new metrics we have defined,
especially the DN, which
accounts for the relative content of the observables in addition to the
absolute one. The results presented for the SDSS
dataset show that the novel DNF method provides the best 
approximation for photometric redshift estimation among the nearest 
neighbour approaches explored. Furthermore, extending the 
tests to the higher redshift datasets of VVDS and PHAT confirms that 
DNF is among the best scored algorithms.

Another quality of DNF is its ability to compute reliable photo-z PDFs which provide a
 useful tool for detecting degeneracies and estimating the redshift distribution, both
 in photo-z bins and for the complete sample. Another key
 advantage of DNF is that the residuals of the fit directly
 provide a reliable photo-z error estimation. In conclusion, DNF can be applied to current and future surveys to
obtain improved cosmological parameters.

\section*{Acknowledgements}
\label{sec:acknowledgements}  

Funding support for this work was provided by the Spanish Ministry of Economy and
Competitiveness (MINECO) through grant FPA2013-47986-C3-2-P and by the Autonomous Community 
of Madrid through the project SPACETEC-CM (S2013/ICE-2822).
Funding for SDSS-III has been provided by the Alfred P. Sloan Foundation, the Participating Institutions, the National Science Foundation, and the U.S. Department of Energy Office of Science. The SDSS-III web site is $http://www.sdss3.org/$.
SDSS-III is managed by the Astrophysical Research Consortium for the Participating Institutions of the SDSS-III Collaboration including the University of Arizona, the Brazilian Participation Group, Brookhaven National Laboratory, Carnegie Mellon University, University of Florida, the French Participation Group, the German Participation Group, Harvard University, the Instituto de Astrofisica de Canarias, the Michigan State/Notre Dame/JINA Participation Group, Johns Hopkins University, Lawrence Berkeley National Laboratory, Max Planck Institute for Astrophysics, Max Planck Institute for Extraterrestrial Physics, New Mexico State University, New York University, Ohio State University, Pennsylvania State University, University of Portsmouth, Princeton University, the Spanish Participation Group, University of Tokyo, University of Utah, Vanderbilt University, University of Virginia, University of Washington, and Yale University.
This research uses data from the VIMOS VLT Deep Survey, obtained from the VVDS database operated by Cesam,
 Laboratoire d'Astrophysique de Marseille, France. 
Also, we thank Hendrik Hildebrandt for the analysis of our data in the context of PHAT project.

\bibliography{dnf}

\end{document}